\documentclass[%
 reprint,showkeys,
 amsmath,amssymb,
 aps,
]{revtex4-2}

\usepackage{graphicx}% Include figure files
\usepackage{dcolumn}% Align table columns on decimal point
\usepackage{bm}% bold math
\usepackage{xcolor}
\usepackage{xr}
\begin{document}

\preprint{APS/123-QED}

\title{Nonresonant optomechanical control of structural phases}% Force line breaks with \\
\thanks{Equal contribution}%

\author{Jiaojian Shi*}
 \email{jiaojian@uw.edu}
 \affiliation{Department of Materials Science and Engineering, Stanford University, Stanford, CA 94305, USA}
 \affiliation{Stanford Institute for Materials and Energy Sciences, SLAC National Accelerator Laboratory, Menlo Park, CA 94025, USA}
 \affiliation{Department of Chemistry, University of Washington, Seattle, WA 98195, USA}

\author{Yijing Huang*}
 \affiliation{Stanford Institute for Materials and Energy Sciences, SLAC National Accelerator Laboratory, Menlo Park, CA 94025, USA}
 \affiliation{Department of Applied Physics, Stanford University, Stanford, CA 94305, USA}
 \affiliation{Stanford PULSE Institute, SLAC National Accelerator Laboratory, Menlo Park, CA 94025, USA}

\author{Christian Heide*}
 \affiliation{Department of Applied Physics, Stanford University, Stanford, CA 94305, USA}
 \affiliation{Stanford PULSE Institute, SLAC National Accelerator Laboratory, Menlo Park, CA 94025, USA}
 \affiliation{Department of Physics, University of Central Florida, Orlando, Florida 32816, USA}
 \affiliation{CREOL, The College of Optics and Photonics, University of Central Florida, Orlando, Florida 32816, USA}
 
 \author{Elias Hildebrand*}
 \affiliation{I. Institute of Physics (IA), Physics of Novel Materials, RWTH Aachen University, Aachen 52056, Germany}
 \affiliation{PGI 10 (Green IT), Forschungszentrum J\"ulich GmbH, 52428 J\"ulich, Germany}

\author{Burak Guzelturk}
 \affiliation{X-Ray Science Division, Argonne National Laboratory, Lemont, Illinois 60439, USA}

\author{Isabel Sedwick}
 \affiliation{Department of Chemistry, University of Washington, Seattle, WA 98195, USA}

\author{Haowei Xu}
 \affiliation{Department of Physics, City University of Hong Kong, Kowloon, Hong Kong SAR, China}
 \affiliation{Department of Nuclear Science and Engineering, Massachusetts Institute of Technology, Cambridge, MA 02139, USA}

\author{Yuejun Shen}
 \affiliation{Department of Materials Science and Engineering, Stanford University, Stanford, CA 94305, USA}
 
\author{Carl Friedrich Sch\"on}
 \affiliation{I. Institute of Physics (IA), Physics of Novel Materials, RWTH Aachen University, Aachen 52056, Germany}
 \affiliation{PGI 10 (Green IT), Forschungszentrum J\"ulich GmbH, 52428 J\"ulich, Germany}

\author{Jan K\"ottgen}
 \affiliation{I. Institute of Physics (IA), Physics of Novel Materials, RWTH Aachen University, Aachen 52056, Germany}
 \affiliation{PGI 10 (Green IT), Forschungszentrum J\"ulich GmbH, 52428 J\"ulich, Germany}

\author{Yuki Kobayashi}
 \affiliation{Department of Applied Physics, Stanford University, Stanford, CA 94305, USA}

\author{Andrew F. May}
 \affiliation{Materials Science and Technology Division, Oak Ridge National Laboratory, Oak Ridge, Tennessee 37831, USA}

\author{Pooja Donthi Reddy}
 \affiliation{Department of Materials Science and Engineering, Stanford University, Stanford, CA 94305, USA}

\author{Viktoryia Shautsova}
 \affiliation{Department of Materials Science and Engineering, Stanford University, Stanford, CA 94305, USA}

\author{Mohammad Taghinejad}
 \affiliation{Department of Materials Science and Engineering, Stanford University, Stanford, CA 94305, USA}

\author{Duan Luo}
 \affiliation{Department of Materials Science and Engineering, Stanford University, Stanford, CA 94305, USA}

\author{Eamonn Hughes}
 \affiliation{Materials Department, University of California Santa Barbara, Santa Barbara, California 93116, USA}

\author{Yukun Li}
 \affiliation{Department of Chemistry, University of Washington, Seattle, WA 98195, USA}

\author{Mark L. Brongersma}
 \affiliation{Department of Materials Science and Engineering, Stanford University, Stanford, CA 94305, USA}

\author{Kunal Mukherjee}
 \affiliation{Department of Materials Science and Engineering, Stanford University, Stanford, CA 94305, USA}

\author{Mariano Trigo}
 \affiliation{Stanford Institute for Materials and Energy Sciences, SLAC National Accelerator Laboratory, Menlo Park, CA 94025, USA}
 \affiliation{Stanford PULSE Institute, SLAC National Accelerator Laboratory, Menlo Park, CA 94025, USA}

\author{Ju Li}
 \affiliation{Department of Nuclear Science and Engineering, Massachusetts Institute of Technology, Cambridge, MA 02139, USA}
 \affiliation{Department of Materials Science and Engineering, Massachusetts Institute of Technology, Cambridge, MA 02139, USA}

\author{Jian Zhou}
 \affiliation{Center for Alloy Innovation and Design, State Key Laboratory for Mechanical Behavior of Materials, Xi'an Jiaotong University, Xi'an 710049, China}

\author{Shambhu Ghimire}
 \affiliation{Stanford PULSE Institute, SLAC National Accelerator Laboratory, Menlo Park, CA 94025, USA}

\author{Matthias Wuttig}
 \affiliation{I. Institute of Physics (IA), Physics of Novel Materials, RWTH Aachen University, Aachen 52056, Germany}
 \affiliation{PGI 10 (Green IT), Forschungszentrum J\"ulich GmbH, 52428 J\"ulich, Germany}

\author{David A. Reis}
 \affiliation{Stanford Institute for Materials and Energy Sciences, SLAC National Accelerator Laboratory, Menlo Park, CA 94025, USA}
 \affiliation{Stanford PULSE Institute, SLAC National Accelerator Laboratory, Menlo Park, CA 94025, USA}

\author{Aaron M. Lindenberg}
 \email{aaronl@stanford.edu}
 \affiliation{Department of Materials Science and Engineering, Stanford University, Stanford, CA 94305, USA}
 \affiliation{Stanford Institute for Materials and Energy Sciences, SLAC National Accelerator Laboratory, Menlo Park, CA 94025, USA}
 \affiliation{Stanford PULSE Institute, SLAC National Accelerator Laboratory, Menlo Park, CA 94025, USA}

\date{\today}% It is always \today, today,

\begin{abstract}
Optical tweezers demonstrate how light can exert forces to trap, repel, and manipulate microscopic particles without absorption. Recent theory has suggested that such forces can extend beyond particle manipulation to drive structural phase transitions in solids. Here we apply this optomechanical principle to tin selenide (SnSe), a material where proximity to several different structural phases gives rise to its high thermoelectric figure of merit and makes it a candidate for a switchable topological crystalline insulator. Whereas the force for standard optical tweezers arises from a gradient in the intensity of a light field, the optomechanical force is mediated by a gradient in the dielectric constant as a function of phonon coordinate. Unlike conventional methods that rely on resonant excitation and absorption through the imaginary part of the dielectric function, this approach operates dispersively through the real part and can be directly driven by Raman processes, enabling selective transitions with reduced energy cost and ultrafast response. Using time-domain Raman scattering, we show that above a critical mid-infrared field strength the $A_g$ Raman modes disappear abruptly without softening, signaling the formation of a new structural phase. This phase, distinct from those induced by heating or carrier excitation, exhibits large-amplitude and long-lived modulations in its optical response. Complementing this observation, we show also evidence for an equivalent DC-field-driven structural phase transformation to a higher symmetry phase, as observed by atom probe tomography. Our study demonstrates the concept of nonresonant optomechanical phase control and defines novel opportunities for synthesizing hidden structural phases with unique functional properties. 

\end{abstract}

%Use showkeys class option if keyword
\maketitle

%\tableofcontents

\section{\label{sec:level1}Introduction}

\begin{figure*}
	\includegraphics[width=\linewidth]{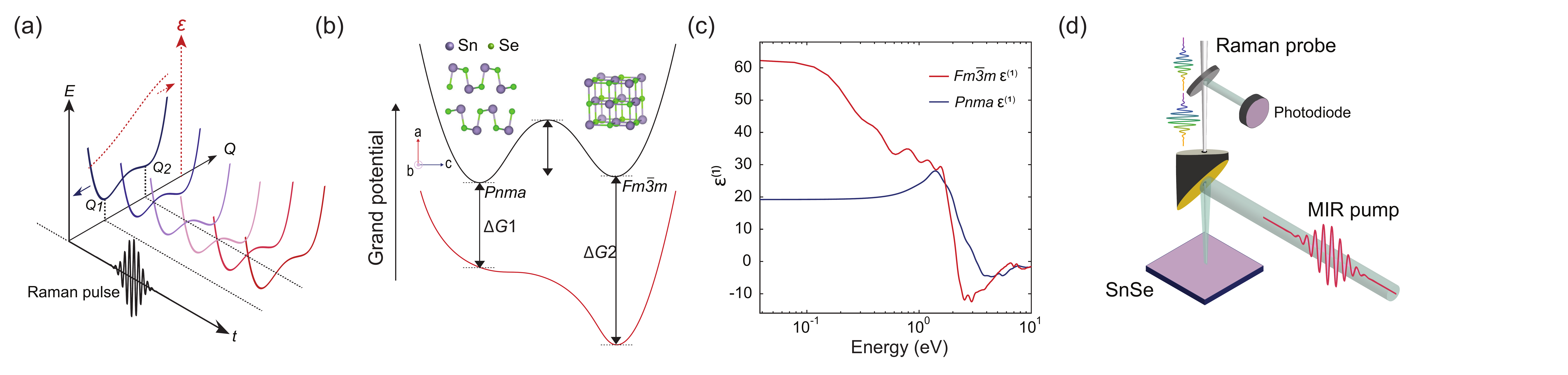}
	\caption{\label{fig:wide}a) The equilibrium asymmetric double-well potential can be modified by laser radiation. Raman pulses interact with a displacement-dependent polarizability $\varepsilon$ that increases from $Q_1$ to $Q_2$, altering the potential surface and favorably lowering the transition barrier to stabilize the final phase at $Q_2$. b) Theoretical calculation of MIR-excited SnSe. Energy curve of intrinsic state (no light exposure, black curve), showing that the $Fm\overline{3}m$ phases are nearly energetically degenerate separated by a barrier. The inset shows the geometric structure of the SnSe compound in the $Pnma$ and $Fm\overline{3}m$ phase. MIR excitation alters the thermodynamic grand potential and favors $Fm\overline{3}m$ over $Pnma$ phase (red curve). c) Calculated real part of the dielectric function of $Pnma$ and $Fm\overline{3}m$-SnSe. d) Schematic illustration of MIR-pump time-domain Raman scattering measurements on SnSe.}
	\label{fig:1}
\end{figure*}

One of the most striking demonstrations of light–matter interaction is the ability of optical fields to exert mechanical forces on matter~\cite{Ashkin1970}. A well-known example is optical tweezers, where a tightly focused laser beam traps or repels microscopic particles depending on the refractive index contrast between the particle and its environment~\cite{Ashkin1986,Block1990}. This dispersive mechanism, widely used in soft matter and biological systems, provides precise and noninvasive control of motion without absorption. Recently, analogous forces have been proposed at the atomic scale, where optical fields act as tweezers for lattice degrees of freedom and both reconfigure and drive motion on the potential energy landscape of molecular systems or solids~\cite{Zhou2018,Zhou2020a}. The proposed driving mechanism suggests that optomechanical forces can be exerted between phases under non-resonant photoexcitation through a  nonlinear Raman-like process~\cite{Shi2025}, as illustrated in Fig.~\ref{fig:1}a. This is distinct from conventional resonant methods, which operate through absorption and typically excite phonons, charges, or spins~\cite{Fausti2011,Sie2019,Nova2019,Li2019,Liu2012,Baum2007,Stojchevska2014,Zhang2016,Stoica2019,Huang2022,Schlauderer2019}. In the energetic description, the free energy of the system is lowered when the light field produces polarization, which subsequently couples to specific atomic coordinates involving multiple normal modes and thermodynamically favors one state over the other, schematically shown in Fig.~\ref{fig:1}b. Though similar in origin to prior impulsive stimulated Raman scattering (ISRS) approaches using visible or near-infrared pulses~\cite{Yan1985,Fahy1994}, the use of mid-infrared  (MIR) pulses far below the bandgap and yet well above any vibrational resonance mitigates laser damage and multiphoton excitation and enables the excitation of non-perturbative atomic displacements. Compared with recent demonstrations of MIR-mediated Raman driving of large atomic displacements and ferroelectric reversal~\cite{Shi2025}, the combination of strong fields at tailored frequencies with large refractive index contrast enables synchronization of multiple phonons, defining novel possibilities for collectively steering the transition toward the desired state.

This optomechanical force has been theoretically predicted to induce large lattice displacements and trigger a variety of phase transitions in different materials, \textit{e.g.}, SnO~\cite{Zhou2018}, SnSe~\cite{Zhou2018,Zhou2020a}, hBN~\cite{Xu2019}, GeSe, SnTe~\cite{Zhou2021a}, MoTe$_2$~\cite{Zhou2021b} and some 1$T'$ TMDs~\cite{Zhou2020b}. Similar phase transformations via susceptibility mismatch have recently been demonstrated with a static electric field~\cite{Lai2021}. The potential realization of this non-resonant mechanism for phase switching may significantly extend the range of phases one can engineer with light, well beyond those demonstrated with resonant gap excitation. This method also poses many advantages over traditional methods.  For conventional resonant excitation approaches~\cite{Forst2011}, anharmonicity leads to shifting of the modes off resonance as the mode amplitude increases, limiting the maximum displacement. On the contrary, laser-driven Raman excitation does not rely on exact resonance tuning but instead exerts a directional force similar to that of optical tweezers, acting directly on molecular or lattice coordinates. Moreover, this non-resonant excitation scheme favors diffusionless phase change with reduced energy consumption and ultrafast switching speed, highly desirable for rewritable nonvolatile memory devices~\cite{Zhou2018}. It also can excite atomic motion along multiple phonon coordinates, which is beneficial in manipulating phases compared with the nonlinear phononics approach~\cite{Forst2011}, especially for those phase transitions where multiple modes are involved~\cite{Wuttig2007,Mujica2003}. 

Among the above-mentioned examples, the group-IV monochalcogenide SnSe is an especially promising candidate~\cite{Zhou2020a}. SnSe has attracted considerable attention as a high-performance thermoelectric material, exhibiting an exceptionally high thermoelectric figure of merit owing to its favorable electronic transport and intrinsically low lattice thermal conductivity~\cite{Zhao2016,Qin2024}. These properties are closely connected to its strongly anharmonic lattice and proximity to multiple competing structural phases. Under ambient conditions, SnSe ordinarily exists in the topologically trivial orthorhombic $Pnma$, but also exhibits a metastable topologically nontrivial rocksalt $Fm\overline{3}m$ phase in epitaxial thin films~\cite{Wang2015,Jin2017}. More specifically, the rocksalt phase is a topological crystalline insulator (TCI) protected by crystal mirror symmetry~\cite{Sun2013}. It has recently been shown that above-bandgap excitation of SnSe can induce a nonthermal lattice distortion towards an orthorhombic distortion of $Fm\overline{3}m$, \textit{i.e.}, $Immm$ phase, but with an amplitude much smaller than the displacement needed for a complete switch to either $Immm$ or $Fm\overline{3}m$ phase~\cite{Huang2022}. Due to the anisotropic optical response of the $Pnma$ phase and the topological band inversion feature of the $Fm\overline{3}m$ phase, there exists a large contrast in the refractive index between these phases. This allows the light to couple primarily to motion along one specific transition coordinate along the modified potential energy landscape and induces lattice instabilities.  Thus, linearly polarized light pulses can theoretically trigger a transition from $Pnma$ to the $Fm\overline{3}m$ TCI phase by lowering its transition barrier, as shown in Fig.~\ref{fig:1}b. Specifically, in bulk SnSe, the small transition barrier from $Pnma$ to $Fm\overline{3}m$ is predicted to vanish when a moderate MIR field (\textit{e.g.}, 5-$\mu$m wavelength with a field strength of $\sim$0.5 V/nm, calculations shown in Fig.~\ref{fig:1}c and Supplementary Fig.~\ref{fig:Dielectrics_overlaid}) is applied along the in-plane armchair direction~\cite{Zhou2020a}. This barrierless transition with below-bandgap light is predicted to be free from conventional nucleation, favoring fast switches to metastable phases on the picosecond timescale~\cite{Zhou2020a}. Therefore, if realized, this novel form of phase control methodology would lead to important applications in nanophotonics, data storage, and quantum devices, allowing for the precise steering of phase transitions at ultrafast speed and reduced energy consumption.

In this work, we study the effect of intense sub-picosecond MIR pulses on the structural and optical properties of SnSe single crystals and thin films. Time-domain Raman spectroscopy was selected as the probe, as group-theoretical analysis predicts that all four $A_g$ Raman-active modes disappear upon transformation of SnSe from the $Pnma$ phase to the rock-salt phase. An abrupt and dramatic suppression of fully symmetric Raman $A_g$ phonon oscillations above a critical field strength is observed, consistent with a structural switch, into a nonequilibrium state. The absence of MIR-induced mode softening contrasts sharply with static heating or near-infrared excitation conditions. Transient reflectivity measurements show long-lived (up to millisecond) and substantial ($>$ 10\%) reflection enhancement, which are consistent with the formation of a metastable phase with dramatically modified index of refractive. All measurements show a strong threshold-like behavior to the MIR field strength. The associated switching energy required is at a 10$^{-3}$-10$^{-4}$ aJ/nm$^3$ level and many orders of magnitude lower than that used in traditional thermally-driven phase transitions~\cite{Rehn2018}. Finally, because the large refractive index contrast between the $Pnma$ and rock-salt phase also exists at the static limit, we show that a DC voltage pulse in conjunction with a short laser pulse in an atom probe can produce the same field-driven phase transition. This transformation is accompanied by a pronounced change in bonding and a transition to a higher symmetry phase, as evidence by atom-probe techniques.

\section{\label{sec:level1}Experimental Approach}

The experimental setup is shown in Fig.~\ref{fig:1}d. MIR pulses at around 5 $\mu$m with a pulse duration of about 100-200 fs were used in our experiments. The photon energy is chosen to be well below the bandgap so that the multiphoton absorption is mitigated~\cite{Yu1981, Chandrasekhar1977}. Additional Fourier-transform infrared measurements in attenuated total reflectance mode (see Supplementary Fig.~\ref{fig:FTIR_SnSe}) confirm the transparency of this frequency range, which is supported by the theory shown in Supplementary Fig.~\ref{fig:Imag_part}. To obtain structural information of the MIR-excited SnSe, we adopt time-domain Raman spectroscopy~\cite{Kuramochi2021} complemented by the standard transient reflectivity measurements. The time-domain Raman spectroscopy is based on the displacive excitation of coherent $A_g$ modes in SnSe with 800-nm pulses~\cite{Weiner1991,Zeiger1992,Huang2022}. and allows us to infer the atomic transformation via the time domain response of associated vibrational modes. The time-domain Raman measurements involve three pulses. A MIR pulse drives the transition, then subsequently, $A_g$ phonon oscillations are excited with an 800-nm pump pulse and read out in the time domain with a time-delayed 800-nm probe pulse. The Fourier transform of the coherent oscillations provides sufficient spectral resolution for resolving $A_g$ modes with good signal-to-noise ratio. The Raman phonon dynamics were captured by scanning the delay of MIR pulses relative to the 800-nm twin pulses. Detailed descriptions can be found in Experimental Section.

\section{Nonlinear Excitation of Multiple Phonon Modes}

We first verified that optomechanical forces can excite multiple phonon modes using MIR-pump transient reflectivity spectroscopy, which involves a MIR pump pulse and an 800-nm probe pulse. Figure~\ref{fig:2}a shows the results of the MIR-pump transient reflectivity experiments at a field strength of 0.48~V/nm, where pronounced oscillations are observed. Fourier analysis of the oscillatory component shows four peaks at $\sim$1, 2, 3.3, and 4.5 THz. They correspond to the known $A_g^{(1)}$, $A_g^{(2)}$, $A_g^{(3)}$, and $A_g^{(4)}$ Raman modes, respectively. Figure~\ref{fig:2}b shows the field dependence of the transient reflectivity dynamics. At low field strengths ($\sim$ 0.2–0.5~V/nm), pronounced phonon oscillations are observed, with amplitudes that increase with field. Above $\sim$ 0.5~V/nm, however, the oscillations become strongly suppressed and nearly vanish at higher fields. In parallel, a huge background signal grows in a highly nonlinear manner. This behavior indicates that MIR excitation drives a structural distortion in SnSe away from equilibrium states into the nonlinear regime.

\section{Structural Transformation Probed by Time-Domain Raman Scattering}

\begin{figure*}
	\includegraphics[width=0.7\linewidth]{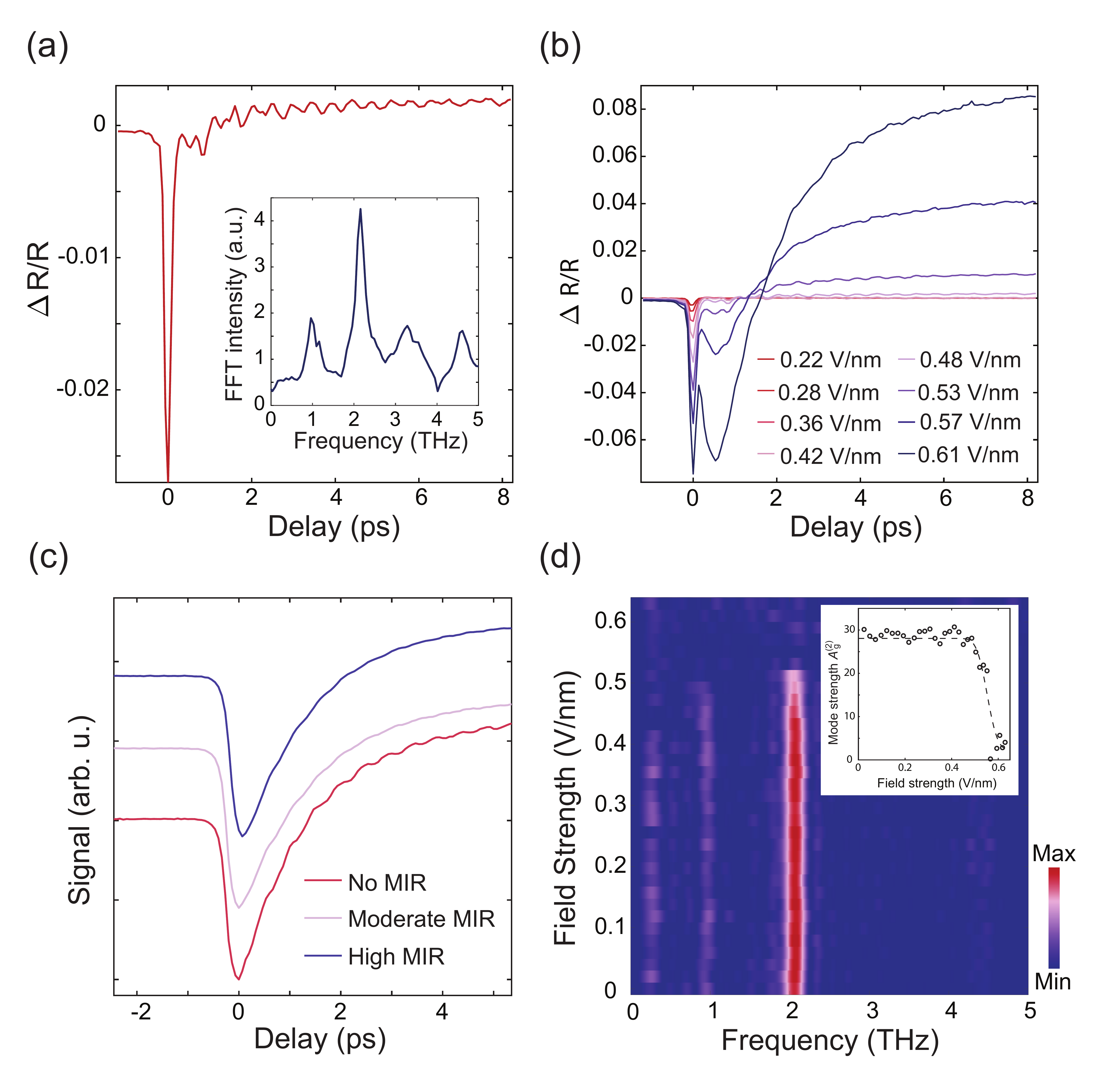}
	\centering
	\caption{\label{fig:wide}a) Transient reflectivity dynamics with a low MIR field strength at 0.48 V/nm. The Fourier transform of the signal's oscillatory part is shown in the inset, and A$_g$ modes are evident. b) Field dependence of transient reflectivity dynamics. At low field strengths ($\sim$ 0.22 – 0.48~V/nm), oscillations are clearly observed and correspond to $A_g$ modes. The phonon oscillations are less pronouced at field strengths higher than $\sim$ 0.53~V/nm. c) 800-nm pump 800-nm probe traces under MIR exposure (wavelength at 5 $\mu$m) at varying field strengths. Coherent $A_g$ phonon oscillations are effectively excited by 800-nm pulses in the absence of MIR exposure while strongly suppressed with MIR excitation. Curves shifted vertically for clarity. d) Fourier transformed reflectivity spectrum at different incident MIR field strengths, measured at 12 ps after MIR arrives. The inset is field dependence scan that shows an abrupt mode suppression above a critical incident field strength.}
	\label{fig:2}
\end{figure*}

To explore the direct linkage between these effects and the structural phase transition, we employed time-domain Raman scattering as a point group symmetry sensitive probe. The Raman measurement is based on the displacive excitation of coherent $A_g$ modes in SnSe and uses a three-beam setup in which a MIR pump pulse is followed by a weak secondary pump and probe at 800 nm to directly probe the phonon response~\cite{Weiner1991,Zeiger1992,Huang2022}. and allows us to infer the atomic transformation via the time-domain response of associated vibrational modes. Figure~\ref{fig:2}c shows the time-dependent reflectivity dynamics excited with 800-nm pulses. Without MIR irradiation, we observe a fast reflectivity decrease around zero pump-probe delay followed by slower recovery dynamics. The oscillatory component of the reflectivity trace corresponds to coherent phonons, and its Fourier transformed spectrum shows three dominant peaks at $\sim$1, 2, and 4.5 THz. They correspond to the known $A_g^{(1)}$, $A_g^{(2)}$, and $A_g^{(4)}$ Raman modes, respectively. Above a moderate MIR field excitation of $\sim$0.6~V/nm (incident fluence $\sim$ 7~mJ/cm$^2$), we observe a substantial reduction of phonon oscillations, which continues to be further suppressed at $\sim$0.65~V/nm. The detailed MIR field dependence scans in Fig.~\ref{fig:2}d show an abrupt suppression of all three $A_g$ modes above the critical field strength. The Raman spectrum evolution as a function of delay times in Supplementary Fig.~\ref{fig:TrRaman_delaydelay} and \ref{fig:Fielddelay_pp} further shows that the mode suppression occurs within a few ps. This represents a signature of a change in point group symmetry, and the observed changes in Raman mode activities resemble the atomic transformation towards $Fm\overline{3}m$ phase, distinct from the high-temperature $Cmcm$ phase as explained below. SnSe undergoes a second-order phase transition to the $Cmcm$ phase at temperatures above 780~K~\cite{Li2015}. The $Pnma$-$Cmcm$ transition mainly involves atomic displacements along the crystallographic $c$ axis toward high symmetry sites. Therefore, the two largely $c$-polarized Raman modes, $A_g^{(1)}$ and $A_g^{(3)}$ become Raman inactive in the $Cmcm$ phase, while the $A_g^{(2)}$ and $A_g^{(4)}$ modes remain Raman-active~\cite{Zhang2015,Liu2018}. In contrast, a  $Pnma$-$Fm\overline{3}m$ or $Pnma$-$Immm$ transition renders all four $A_g$ modes Raman inactive, as in the higher symmetry $Fm\overline{3}m$ or $Immm$ phase, all the atoms take high symmetry position, as detailed in Supplementary Note~\ref{sn:symmetry_analysis}.

Below the critical field strength, no apparent change in the frequency of $A_g^{(2)}$ (Fig.~\ref{fig:2}b) and $A_g^{(1)}$ (Supplementary Fig.~\ref{fig:Fdep_Ag1}) is observed under increased MIR field strengths, which is in sharp contrast with the apparent phonon softening upon heating to higher temperatures, as shown in Supplementary Fig.~\ref{fig:Temp_dep}. The softening of $A_g^{(2)}$ mode is evident as the temperatures increase towards the $Cmcm$ phase, which is often expected in a second-order phase transition~\cite{Shirane1969,Landau1980,Dove1997,Aranson2002}. The time-domain traces at varying temperatures are shown in Supplementary Fig.~\ref{fig:Temp_dep_details}. We note that above-gap excitation also induces $A_g$ mode softening~\cite{Huang2022,Han2022}, akin to the second-order phase transition. Clear phonon softening is also observed for near-infrared (NIR) excitation at a 2-$\mu$m wavelength, as shown in Supplementary Fig.~\ref{fig:NIR_fluence}, which is likely due to two-photon absorption and subsequent heating effects. The distinction of MIR-driven behaviors from those second-order transitions rules out the transition toward $Immm$ phase via NIR excitation. Such an observation is expected from the optomechanical driving mechanism. The grand free-energy landscapes are only deformed when MIR is present, while in the carrier/heating cases, external perturbations persist and still act on the potential surface after the pump, hence modifying the phonon frequencies~\cite{Huang2022}.

\section{Giant MIR-Driven Modification in Optical Properties}

\begin{figure*}
	\includegraphics[width=\linewidth]{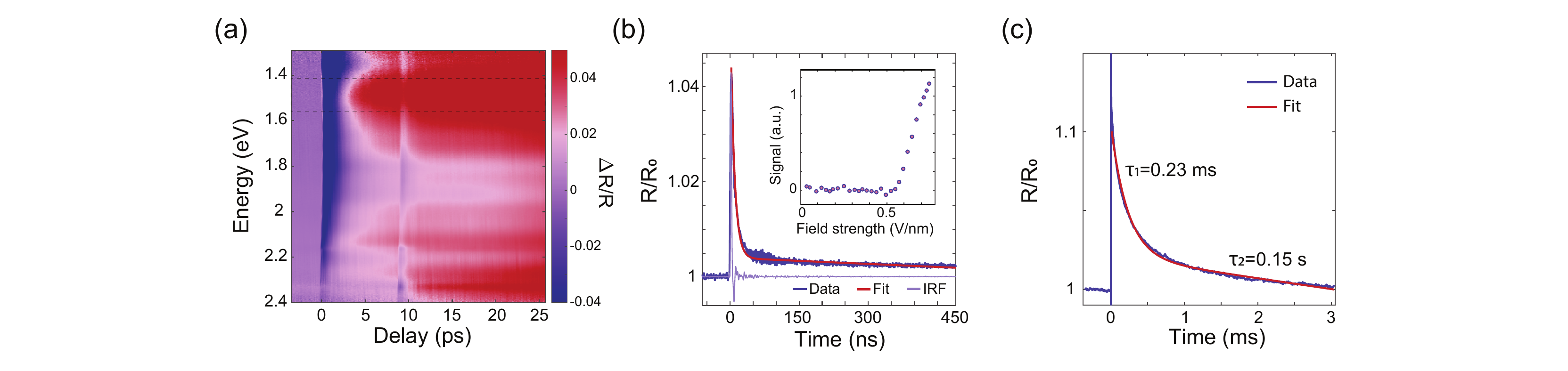}
	\caption{\label{fig:wide}a) Transient reflectivity measurement shows varying responses at different probing wavelengths. Around 8 ps, a copy of the MIR laser pulse is found, which again decreases the reflectivity. A steady-state reflectivity enhancement emerges and develops upon MIR excitation with no noticeable decay up to 25 ps after the excitation. b) Transient reflectivity dynamics at 633~nm measured with an oscilloscope show biexponential decay with long-lived responses potentially indicative of the induction of a metastable phase. IRF: instrument response function. The inset shows that the amplitude of the fast decay signal show a threshold-like field dependence. c) Transient reflectivity dynamics at 1520~nm measured with an oscilloscope show a dramatic reflectivity increase around 10\% at prolonged ms timescales.}
	\label{fig:3}
\end{figure*}

We further study the optical properties of the MIR-induced transient phase. Figure~\ref{fig:3}a shows a long-lived signal after MIR excitation with different optical properties across a broad spectral range from 1.3 to 2.4 eV. When summed over 1.4-1.55 eV spectral range, a steady-state reflectivity enhancement of about 7\% emerges and shows no noticeable decay over 25 ps. Figure~\ref{fig:3}b shows the MIR-induced dynamics probed at 633-nm wavelength measured with an oscilloscope and can be fitted with double exponentials with a fast ($\tau_1 = 8$ ns) and a slow decaying ($\tau_2 = 0.24$ ms) component. The field dependence of reflectivity enhancement in Fig.~\ref{fig:3}b shows the steady-state signals appear promptly above a threshold and with amplitudes that increase in a highly nonlinear fashion, signaling the onset of a phase switch. The lifetime also tends to increase with higher field strengths (see Supplementary Fig.~\ref{fig:Fdep_analysis}), consistent with the picture of a more stabilized non-equilibrium phase induced by stronger MIR pulses, thereby extending the recovery time. Figure~\ref{fig:3}c shows the responses are greatly strengthened by probing the dynamics at 1520-nm wavelength closer to the bandgap~\cite{Yu1981,Chandrasekhar1977}. The reflectivity boost is as big as 10\% and lasts for ms timescales, sufficiently large and long for device applications. Such a long-lived response is also distinct from the carrier relaxation dynamics~\cite{Ye2019,Yin2021} and indicative of the induction of a metastable phase. We also found the sample almost fully recovered in a fatigue testing measurement after 10$^5$ MIR excitation cycles (see Supplementary Fig.~\ref{fig:Fatigue}), which is consistent with a displacive type of structural distortion.

\section{Significant Bonding and Symmetry Changes Under DC Fields}

To further validate the optomechanical mechanism, we adopted DC-field driving as the refractive index of the rock-salt phase is much greater than that of the $Pnma$ phase~\cite{Flitcroft2022}, similar to the MIR case. Therefore, a similar phase transition is expected upon DC field driving. Indeed, we can also produce a related phase in atom probe tomography (APT). In this experiment a sharp tip is either exposed to a strong voltage pulse (voltage mode) or a slightly lower DC voltage in conjunction with a short laser pulse (laser-assisted mode). If a sufficiently high voltage is applied, ionic fragments are dislodged from the tip, which are characterized in terms of their flight time and their position where they hit the detector. This enables the determination of the mass to charge ratio of the ion. The position on the detector allows a reconstruction of the fragment’s position in the tip before field evaporation. APT hence creates an image of the elemental distribution in a tip with near atomic resolution. Yet, APT provides another crucial piece of information. Two quantities have been shown to help discriminate different mechanisms of bond rupture, which are apparently related to differences in chemical bonding. These two quantities are the probability of forming molecular ions (PMI), instead of dislodging single atoms and the probability of multiple events (PME). The latter quantity describes the probability that more than one ion is dislodged simultaneously, \textit{i.e.}, upon one pulse~\cite{Cojocaru2024}. 

\begin{figure*}
	\includegraphics[width=0.6\linewidth]{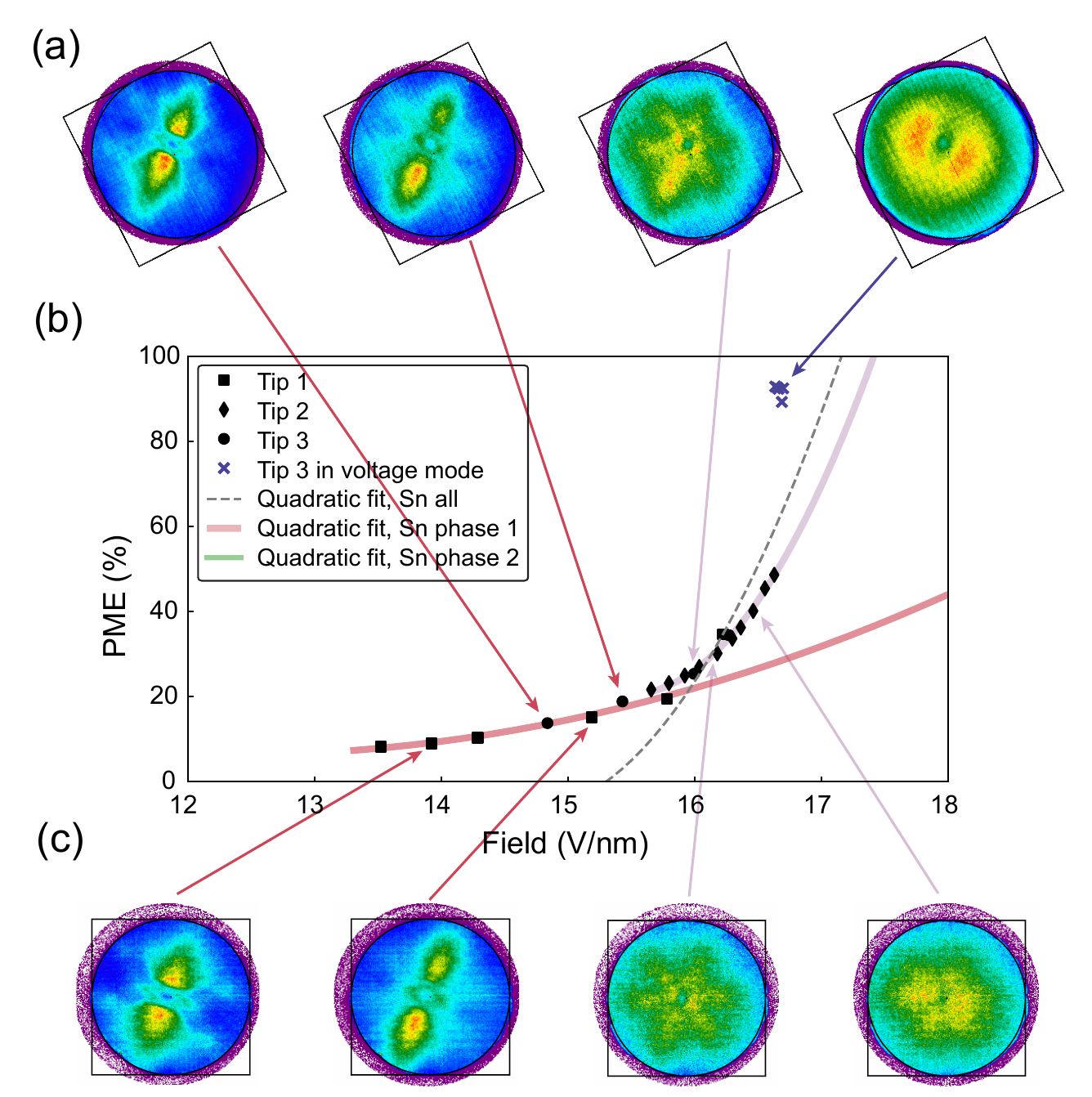}
	\centering
	\caption{\label{fig:wide}Atom probe tomography under DC fields. a) and c) Detector event histograms at different electrical field strengths, as indicated by the arrows pointing onto the associated field strength and PME pair. The histograms in (a) are rotated about 120 degrees for better comparability with the histograms in (c) as tip 3 was taken from another lift out (site on the sample) than tip 1 and tip 2. On the outmost right in (a), a histogram from a measurement in voltage mode is shown. b) PME from different measurements is plotted against the electrical field at which the measurement was taken. Three different tips were measured for this and fitted afterwards with a parabolic fit. Fitting datapoints with two different fit functions (as given by the fits in red and green with both R$^2$  = 0.98) yields a much better fit quality, than a single fit over all points, as given by the dotted grey line (with R$^2$ = 0.69). The measurements in voltage mode were excluded from the fits, because the evaporation mechanism is different to the laser assisted evaporation.}
	\label{fig:4}
\end{figure*}

Interestingly, for SnSe we find two distinctively different bond ruptures. For small applied voltages, we observe a low PME, characteristic for covalent solids, while for large applied voltages and hence high applied electric fields, a significantly higher PME is detected. This indicates that the bond rupture and hence bonding in both phases differs. This is further supported in Fig.~\ref{fig:4}, where the PME and the corresponding detector histogram is depicted for different applied fields. For a large number of solids, a nonlinear dependence between PME and the field applied to the tip has been seen, which can be best fit by a parabolic dependence. Clearly, the measured data cannot be accounted for by a single parabolic dependence. Instead, the data can be described much more convincingly with two different parabolas, evidence for two different crystalline states of the sample upon increasing electric field. Apparently, above a critical field strength of about 15.8~V/nm, a switch to a new phase is encountered which is characterized by another field dependence of the PME. This phase is characterized by a significantly larger PME, and a different detector histogram. The latter can be attributed to differences in atomic arrangement. At low electric field, the detector histogram has a rather low symmetry, in line with the low symmetry we expect from a sample with orthorhombic ($Pnma$) symmetry. At high electric field, however, a much more symmetric detector histogram is observed, which can be attributed to a cubic phase. 

This detector event histogram is a stereographic projection of the surface of the APT-tip investigated~\cite{Khan2016}. A change in the symmetry of the crystal therefore directly changes the symmetry of the resulting histogram. Changes from a twofold to a sixfold symmetry in the histograms, thus, are indicative for a change to a higher symmetry, as the orientation of the crystal inside the tip remains the same. Simultaneously, however, the strong increase in the PME lowers the resolution of the detector histogram. This effect is particularly pronounced in voltage mode, when the PME exceeds 90\%, resulting in the inability to resolve any other symmetry apart from the pole in the middle of the histogram under those conditions. In laser-assisted mode, \textit{e.g.}, at a field strength of 16~V/nm, a sixfold symmetry is discernible, consistent with a cubic symmetry. Further discussion is included in Experimental Section and Supplementary Note~\ref{sn:DC_APT_discussion}.

\section{Theoretical Modeling of the Structural Phase Transition}

\begin{figure*}
	\includegraphics[width=\linewidth]{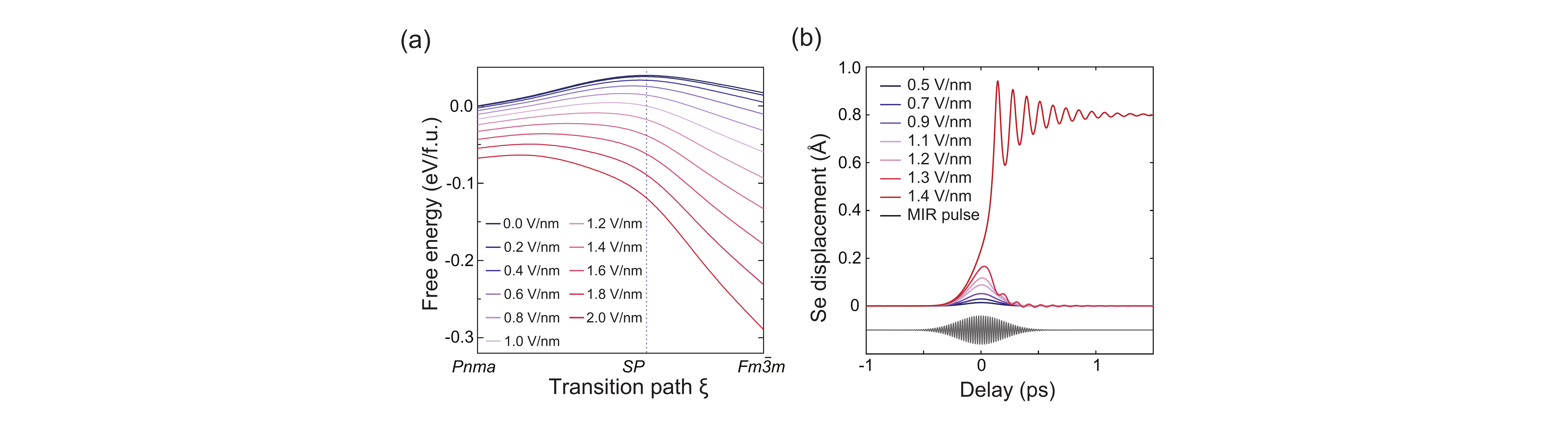}
	\caption{\label{fig:wide}a) Grand potential of SnSe per formula unit under varying MIR fields. The energy is shifted with respect to the intrinsic (no MIR irradiation) $Pnma$ phase. SP: saddle point. b) Time-dependent simulation of Se-atom displacement dynamics driven by MIR fields at increasing field strengths. The black curve shows the MIR shape used in the simulation and offset for clarity.}
	\label{fig:5}
\end{figure*}

We first carried out calculations to establish the theoretical basis of this new optomechanical force in SnSe. Based on the calculated dielectric functions of $Pnma$ and $Fm\overline{3}m$ phases, and the intermediate states along the transition pathway, the interaction with the MIR fields leads to an additional term in the free energy density, $G_F(\xi,t) =U(\xi)-\frac12\vec{F^*}(\omega_0,t)\cdot\vec{\varepsilon^{(1)}(\omega_0;\xi)}\cdot\vec{F}(\omega_0,t)V_p$, given by the thermodynamic theory and shown in Fig.~\ref{fig:5}a. Here, $\xi$ labels the generalized coordinates along the transition path, $U(\xi)$ is the equilibrium potential in the absence of external fields, $\varepsilon^{(1)}(\omega_0;\xi)$ is the real part of the dielectric function tensor of state $\xi$ at MIR frequency $\omega_0$, and $\vec{F}$ is the incident MIR electric field applied, and $V_p$ is the primitive unit cell volume. $G_F(\xi,t)$ modifies the energy landscape with varying MIR field strength $F$. As shown in Fig.~\ref{fig:5}a, the $Fm\overline{3}m$ phase becomes energetically more stable than the pristine $Pnma$ phase if a MIR field $F \gtrsim$ 0.5~V/nm ($\sim$5~mJ/cm$^2$) is applied. Moreover, the transition barrier between the $Pnma$ and $Fm\overline{3}m$ phases decreases with increasing $F$. When $F$ exceeds $\sim$1.1~V/nm, the barrier vanishes, and the system can transit to $Fm\overline{3}m$ phase barrierlessly. 

To provide a time-resolved view of MIR interactions with SnSe, we adopted classical equation of motion given by $m\frac{d^2 \xi}{dt^2}=-\frac{\partial G_F}{\partial \xi}=-\frac{\partial U}{\partial \xi}+\frac{1}{2}\frac{\partial \varepsilon}{\partial \xi} F^*(t)F(t)V_p$. By parametrizing the $G_F(\xi,t)$ in Fig.~\ref{fig:5}a, the time-dependent atomic displacements can be obtained by integrating equations of motion, as shown in Fig.~\ref{fig:5}b (details are included in Experimental Section). Below 1.2~V/nm, the Se displacement increases during the presence of MIR fields and returns to its original value after the excitation, which transits to a new value corresponding to the $Fm\overline{3}m$ phase at 1.4~V/cm. Such a level of field strengths correspond to an effective optomechanical force along the transition path given by $\frac{1}{2}\frac{\Delta \varepsilon}{\Delta \xi}F^*(t)F(t)V_p$ which is approximately 1 nanonewton for reasonable values of $\delta\varepsilon / \delta\xi$, thus more substantial than typical forces used in optical tweezers~\cite{Jannasch2012}. We also note that this model clarifies the connection of the optomechanical mechanism with ISRS. If the potential $U(\xi)$ can be simplified to a parabola, \textit{i.e.}, $U(\xi)=\frac12\Omega^2\xi^2$ and $\Omega$ is the parabolic curvature at $\xi=0$, this yields the formulation for ISRS in the perturbative limit~\cite{Merlin1997}.

\section{Mechanistic Discussion and Technological Significance}

To assess its technological relevance, we further calculated the energy consumption in the non-resonant photoexcitation regime and compared it with other established or emergent phase switch schemes. The energy density applied is given by $\frac12\varepsilon_0|F|^2$ , which gives about $3\times10^{-3}$ aJ/nm$^3$. By multiplying the unit cell volume of the $Pnma$ structure, this value corresponds to $\sim$ 0.5 meV/atom or 0.004 eV/unit cell. We note that our estimates assume both negligible changes in the optical properties during the MIR pulse and that the phase-transition timescale is longer than the 100-200 fs MIR pulse duration. Compared with traditional thermally-driven phase changes in Ge$_2$Sb$_2$Te$_5$ (GST), such an energy density at the order of 10$^{-2}$ aJ/nm$^3$ is 10$^5$-10$^7$ times lower than the experimentally measured values in GST~\cite{Lacaita2013} and lower than the theoretical adiabatic lower limit in GST~\cite{Rehn2018}. The dramatic discrepancy in the experimental and theoretical energy consumption of GST is due to challenging heat management with waste heat flow out of the material~\cite{Rehn2018}. This is in stark contrast with this non-resonant photoexcitation scheme, where heating is not required for phase engineering. We further estimate the switching efficiency (energy density divided by the volume of the material switched) to be at a 10$^{-3}$-10$^{-4}$~aJ/nm$^3$ level and comparable or lower than other light-induced phase transitions summarized in Supplementary Note~\ref{sn:literature_comparison} and Table~\ref{table:Energy_comparison}.

\section{Conclusion}

In conclusion, we demonstrate a novel non-resonant means for controlling materials structure with combined time-domain Raman scattering and transient reflectivity measurements. The significant suppression of $A_g$ Raman modes by MIR pulses is consistent with the distortion of a new structural phase, and the absence of discernible Raman mode softening is distinct from the sample responses at evaluated temperature or under NIR excitation. The concomitant reflectivity enhancement over up-to-millisecond lifetime corroborates the theoretical simulation of a long-lived metastable phase induction. The giant reflectivity increase up to about 10\% further suggests applications in phase-switch devices that exploit this long-lived nonequilibrium state and the associated changes in optoelectronic properties. This mechanism outperforms existing alternatives in terms of low energy consumption and opens up tantalizing perspectives in the search for novel phases of matter inaccessible from the conventional resonant or above-bandgap excitation~\cite{Huang2022} scheme. Complementary DC-field APT measurements reveal an analogous field-driven transformation with pronounced bonding changes, extending this optomechanical control beyond optical excitation. With further tailored excitation (\textit{e.g.}, pulse sequences, longer MIR pulses with frequency tunability, or DC assistance), such non-resonant approach could stabilize exotic states with profound implications in the ongoing research on light-induced topological phase transitions~\cite{Sie2019,Mclver2020}.

% Experimental section

\section{Experimental Section}
\subsection{\label{sec:level2}SnSe sample synthesis} 
Samples consist of bulk single-crystal SnSe grown with a Bridgman-type technique~\cite{Li2015}, polished with [100] surface normal for optical experiments, and epitaxial thin films grown in a Riber Compact 21 MBE system on (001)-oriented GaAs substrates, which were prepared by a GaAs regrowth and arsenic capping process. The substrate was held at $400^\circ$C for 10 minutes to desorb the arsenic cap. After which, the substrate was exposed to a compound PbSe flux for 35 s at $400^\circ$C to modify the substrate surface and seed growth of a (001)-orient film. The substrate was then cooled to $330^\circ$C for 30 s and then $280^\circ$C for 50 s while exposed to the compound PbSe flux to nucleate and then grow approximately  5 nm of (001) PbSe. Following these surface preparation steps, the sample was exposed to a compound SnSe flux for 16 minutes at $280^\circ$C to deposit approximately 50 nm of SnSe.\\
\subsection{\label{sec:level2}Ultrafast mid-infrared and near-infrared pulse generation}
MIR pulses were generated in a 0.5 mm thick GaSe crystal by difference-frequency mixing of the signal and idler outputs of a high-energy optical parametric amplifier (OPA). The OPA was pumped with $\sim$ 35 fs pulses from a commercial Ti:Sapphire regenerative amplifier (800 nm central wavelength 800 nm, 6 mJ) and has an output power at signal and idler at 1.2 mJ and 1 mJ. The generated MIR pulses from the GaSe crystal were first filtered by MIR filters (Thorlabs FB5000-500 and Edmund Optics \#62-643) and then collected and expanded by a pair of 2-inch diameter parabolic mirrors before being tightly focused to the sample by a third parabola with a 2-inch diameter and 2-inch effective focal length. The MIR field strengths were controlled by a half-wave plate and polarizer (Thorlabs WP25H-K). NIR pulses at 2-$\mu$m wavelength were generated from a high-energy OPA and filtered by NIR filters (Thorlabs FELH1100). The beam was attenuated with a half-wave plate and polarizer (Thorlabs AHWP10M-1600 and WP25M-UB) before focusing on the sample by parabolic mirrors.\\
\subsection{\label{sec:level2}Time-domain Raman scattering} For the time-domain Raman scattering probe, the setup is similar to the previous work~\cite{Huang2022}. The 800-nm pump pulse was chopped at 0.5 kHz and excited bulk SnSe crystal with a fluence of $\sim$ 0.5 mJ/cm$^2$, and the 800-nm probe beam was detected by a photodiode and a lock-in amplifier. A reference photodiode was used to suppress noise. Details are provided in Supplementary Note~\ref{sn:setup_details} and Fig.~\ref{fig:TrRaman_setup}. The high-temperature experiment was performed using a heating stage (INSTEC) in a vacuum ($\sim$ $10^{-2}$ mbar) on the SnSe thin-film sample.\\
\subsection{\label{sec:level2}Transient reflectivity} For the transient reflectivity probe, the reflected white light from bulk SnSe sample was spectrally resolved by a spectrometer (Ocean Insight, OCEAN-FX-VIS-NIR). To obtain the change in reflectivity ($\Delta R/R$) induced by each MIR pulse, we synchronously chopped the MIR pump beam at 0.5 kHz, thereby blocking every other pump beam. The acquired data were subsequently corrected to account for the white-light chirp~\cite{Pein2019}. The copy of the MIR pulse at around 8 ps in Fig.~\ref{fig:3}a is likely due to reflections in the optical path and not from the pump reflection in the sample. For the transient reflectivity probe with an oscilloscope readout, we encoded the reflectivity change into the intensity of continuous-wave He-Ne laser light at 633 nm. The reflected laser light is coupled to a fast amplified photoreceiver (Thorlabs, PDA10A2). The electronic signal is collected by a 6-GHz oscilloscope (Agilent, DSA90604A).\\
\subsection{\label{sec:level2}APT setup} For APT, a CAMECA LEAP 5000XS with a laser wavelength of 355 nm is employed. Measurements have been performed at 40 K, with an average detection rate of 1.0\% and a laser pulse repetition rate of 200 kHz. Pulse energies were changed from 2 up to 19 pJ in order to change the effective evaporation field. Data reconstruction was performed using APSuite 6.3 software and the EPOSA analysis package. From the reconstructed datasets, the PMI and PME were extracted to characterize bond rupture during field evaporation. In order to investigate the influence of the electric field on the material, the evaporation field had to be changed. This was achieved by applying different laser pulse energies. Since the experiments are performed for constant average detection rate, increasing the laser pulse power reduces the voltage applied to the tip. For SnSe, we observe a phase transition upon increasing field strength (and simultaneously decreasing laser pulse power). It is thus the applied field which is responsible for the phase transition.\\
\subsection{\label{sec:level2}Theoretical calculation details} The first-principles calculations in this work are based on density functional theory (DFT)~\cite{Hohenberg1964,Kohn1965} as implemented in Vienna ab initio simulation package (VASP)~\cite{Kresse1996,Kresse1996a}. Core and valence electrons are treated by projector augmented wave (PAW) method~\cite{Blochl1994} and a plane wave basis set with 300 eV cutoff energy, respectively. The hybrid functional proposed by Heyd, Scuseria, and Ernzerhof (HSE)~\cite{Heyd2003} is adopted. The
first Brillouin zone is sampled by a $\Gamma$-point-centered $k$-mesh with a grid density of $2\pi \times 0.02~$\AA$^{-1}$ along each dimension. The time-dependent simulation is done by numerically solving the equation of motion to get $\xi(t)$, given the initial condition that $\frac{d\xi}{dt}=0$ and $\xi$ is in the $Pnma$ phase. We parametrize $G_F(\xi,t)$ by interpolating $U(\xi)$ and $\varepsilon(\xi)$ from the first-principle calculation and fitting with a fourth-order polynomial function (details are included in Supplementary Note~\ref{sn:displacement_simulation}).\\

\begin{acknowledgments}
This work was supported by the US Department of Energy (DOE), Office of Basic Energy Sciences, Division of Materials Sciences and Engineering, under contract no. DE-AC02-76SF00515. I.S., Y.L., J.S. acknowledge the start-up funding provided by the University of Washington. H.X., J.L. acknowledge support by an Office of Naval Research MURI through grant \#N00014-17-1-2661. M.W. acknowledges financial support from the Federal Ministry of Research, Technology and Space (BMFTR, Germany) in the project NEUROTEC-II (project numbers: 16ME0398K) as well as support by the Deutsche Forschungsgemeinschaft within SFB 917 (Nanoswitches) and computing time provided by the NHR Center NHR4CES at RWTH Aachen University (project number p0020357). J.Z. acknowledges the support by the National Natural Science Foundation of China under grant No. 21903063. C.H. acknowledges support from the U.S. Department of Energy, Office of Science through the AMOS program and the Alexander von Humboldt Research Fellowship. S.G. acknowledges support from the U.S. Department of Energy, Office of Science through the AMOS program. Bulk crystal synthesis (A.F.M.) was supported by the U. S. Department of Energy, Office of Science, Basic Energy Sciences, Materials Sciences and Engineering Division.
\end{acknowledgments}


\providecommand{\noopsort}[1]{}\providecommand{\singleletter}[1]{#1}%

\end{document}

\preprint{APS/123-QED}
	
	\title{Supplementary Information\\Nonresonant optomechanical control of structural phases }% Force line breaks with \\
	\thanks{A footnote to the article title}%

	\date{\today}% It is always \today, today,
	\maketitle

\section{Details of MIR-pump time-domain Raman scattering probe\label{sn:setup_details}} 
 
\begin{figure*}[htb!]
	\includegraphics[width=\textwidth]{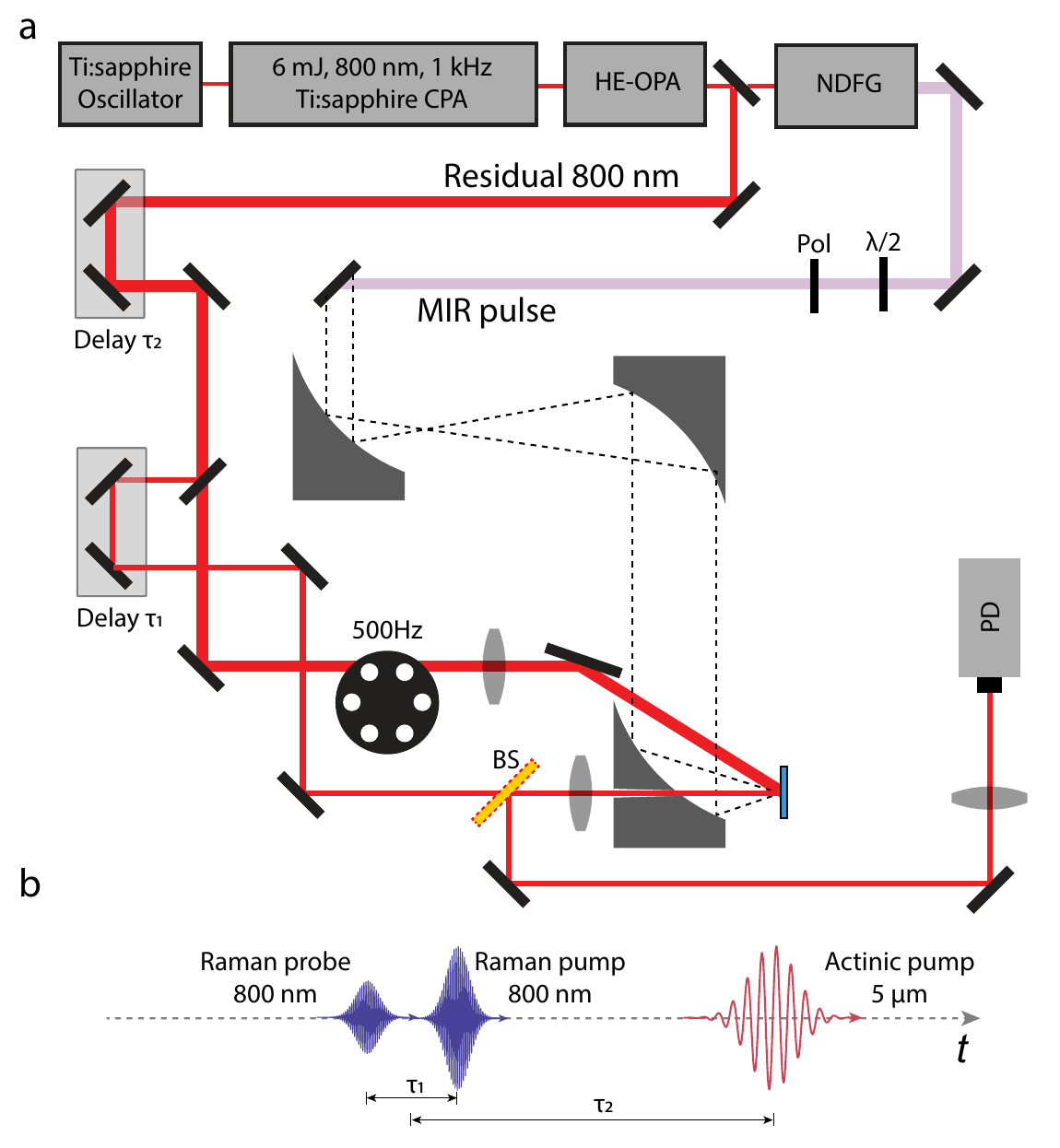}
	\centering
	\caption{\textbf{Schematic illustration of the setup used for MIR-pump time-domain Raman scattering measurements.} a) MIR excitation at 5 $\mu$m is generated from a non-collinear difference frequency generator and focused on the sample with a parabolic mirror imaging system. The time-domain Raman scattering probe is realized with the residual 800-nm beam from the HE-OPA. HE-OPA: high-energy optical parametric amplifier. NDFG: non-collinear difference frequency generator. Pol: polarizer. $\lambda/2$: half waveplate. BS: beamsplitter. PD: photodiode. b) The pulse train sequence used in the experiment.} 
	\label{fig:TrRaman_setup}
\end{figure*}

Supplementary Fig.~\ref{fig:TrRaman_setup} shows the schematics of the MIR-pump time-domain Raman scattering probe setup. The power of the MIR actinic pulse was controlled with a MIR half-wave plate and polarizer and focused tightly on the sample with parabolic mirrors. Time-domain Raman scattering consists of twin 800-nm pulses that serve as the Raman pump and Raman probe pulses. By scanning the relative delay ($\tau_1$) between Raman pump and Raman probe pulses, the time-dependent reflectivity dynamics can be obtained in the first order. By subtracting the electronic background induced by the 800-nm Raman pump pulse, coherent phonon oscillation can be extracted, and its Fourier transformed spectrum contains Raman phonon fingerprints. By scanning the relative time delay between the actinic MIR pump and 800-nm twin pulses ($\tau_2$), the phonon dynamics induced by MIR excitation can be acquired. In this measurement, the time resolution is mainly limited by the width of the scanning window of the Raman pump and probe pulses, which is typically around 5 ps to yield Raman spectra with sufficient spectral resolution. The pulse sequence is given in Supplementary Fig.~\ref{fig:TrRaman_setup}b. This time-domain method can cover low-frequency phonons that are hard to access with the frequency-domain Raman method (\textit{e.g.}, spontaneous Raman, femtosecond stimulated Raman scattering). We also note that the time-domain Raman scattering measurements are constrained by frequency-time uncertainty. Specifically, to monitor the low-frequency $A_g$ modes with terahertz resolution, the temporal resolution of the probe is about 1 ps or worse. Such probe temporal resolution does not support measurements of near time-zero responses of an ultrafast MIR pulse with a duration of about 100-200 fs in our case.

\section{Symmetry change upon phase transition\label{sn:symmetry_analysis}}
The structural phase transition from $Pnma$ to $Cmcm$ is concurrent with the freezing of the $z_s$ degree of freedom of atoms when transformed into high symmetry positions (see Supplementary Table~\ref{table:Wyckoff}). Thus it is expected that the two Raman modes mainly polarized along the $c$ axis disappear in the $Cmcm$ phase, while modes mainly polarized along the $a$-axis remain Raman active. In contrast, the $Pnma$-$Fm\overline{3}m$ transition will freeze both $x_s$ and $z_s$ degrees of freedom when transformed into high symmetry positions, and all four Raman active modes are expected to disappear. We further show the temperature-dependent time-resolved Raman where the $A_g^{(2)}$ modes do not disappear as it approaches the $Pnma$-$Cmcm$ phase transition, consistent with literature~\cite{Zhang2015}.

\section{Energy density and lifetime comparison of light-induced transient states\label{sn:literature_comparison}}
Supplementary Table~\ref{table:Energy_comparison} shows the energy density and lifetime of transient states in this work compared to other reports on light-induced phase transitions in terms of lattice, electronic, and spin degrees of freedom. We quote two modes of computing the energy density. (1) free-space incident energy density $E_v$ given by the $E_v = F/ct$, where $F$ is the excitation fluence, $c$ is speed of light, and $t$ is the laser pulse duration. This parameter is essentially the applied energy density and is independent of the properties of the material. (2) Material switching efficiency (heat load) $E_s$ given by $E_s = F/d$, where $d$ is the penetration depth. This value characterizes the energy applied divided by the volume of the material switched, which depends on the optical properties of the material. SnSe is theoretically transparent to MIR~\cite{Zhou2020a} and a very long penetration depth ($>$ 1 mm) is found experimentally~\cite{Das2015}. The actual penetration could deviate from the values above due to the highly nonlinear MIR-driven responses in SnSe. Thus we performed fluence dependence measurements of the transmitted MIR power through the sample with a thickness of about 200 $\mu$m, as shown in Supplementary Fig.~\ref{fig:Incident_transmission}. About 40\% of MIR is transmitted through the sample at low fluences, which is expected since MIR is well below the bandgap. However, we did not measure the reflected power due to technical challenges from sample surface inhomogeneity; thus, the transmitted power can be used to estimate the upper limit of the absorption. Although there are some changes in the transmission above the transition threshold, these measurements indicate the penetration depth of SnSe under strong MIR excitation is larger than 100 $\mu$m. Despite that, since the optomechanical mechanism for excitation is distinct from heating and carrier excitation, the dissipative energy cost is reduced, and the switching efficiency is maximized. 

We also would like to point out its practical device applications. Although the energy budget of such MIR laser sources are currently high, there is an increasing availability of low-cost, high-quality on-chip MIR sources and MIR waveguides~\cite{Yao2012,Schadle2016}. Moreover, the burgeoning field of MIR metamaterials~\cite{Watts2012,Liu2018} holds great promise for augmenting and customizing the response, including pulse duration and frequency modulation, thus rendering them suitable for miniaturized phase switching applications. Consequently, we posit that the realization of practical implications is conceivable.

\section{Alternative scenarios behind the MIR-driven dynamics\label{sn:alternative_scenarios}}
For the possible alternative mechanism of carrier generation via MIR excitation, it is known that photons with energy above or slightly below the optical band gap energy can efficiently generate carriers through multiphoton absorption. Photons with very low energy, such as terahertz photons, behave classically when interacting with materials and exhibit Zener tunneling and/or impact ionization. In this regime, longer pulses, \textit{i.e.}, lower photon energies, are more favorable for carrier generation because it allows more extended time for carrier tunneling and multiplication. In our study, we utilized intermediate photon energy levels of approximately 250 meV, a range in which Zener tunneling/impact ionization is disadvantageous and requires more than five-photon absorptions to achieve the bandgap. Therefore, any potential carrier excitation likely does not play a dominant role, as demonstrated by the transmission measurements described above. The observation of distinctly different Raman mode activity changes further supports this notion. 

We also considered the possibility of strain-induced phase transitions and trivial optical nonlinearities. Although strain is produced as a byproduct of the MIR-induced phase changes, the rapid transition rate on the picosecond timescale and the observed changes in the structural factor suggest that strain alone is not responsible for the sample switch. Moreover, since the sample is pumped in transparency by MIR fields, the optomechanical force appears to be the only known mechanism capable of generating such strain that we are aware. With regard to the possibility of trivial optical nonlinearities as the cause, all of our observed phenomena occurred picoseconds after the MIR left the sample, indicating that there were no optical nonlinearities between MIR and probe fields involved in the observation. 

For the MIR-induced phase transition dynamics, several pieces of evidence shows the presence of strain. However, we have reservations that simply straining the sample or introducing disorder between unit cells would not induce changes to the Raman peak intensity, as those observations indicate transformation within the unit cell. Additional strain responses could also arise from electrostrictive responses, although the applied fields in our experiments are only on instantaneously at time zero. 

\section{Theoretical calculations of Raman force directions in SnSe\label{sn:Raman_signs}}

To calculate the Raman force, we first calculate the phonon normal modes $\bm{r}_n$, where $n$ is the index of phonon modes at the $\Gamma$ point. Then, we displace the atoms along each phonon mode, and calculate the corresponding dielectric tensor $\varepsilon$. The Raman force along the $n$-th phonon mode is obtained by 
\begin{equation}\label{eq:Raman force}
	\bm{F}_n \propto \frac{\partial \varepsilon}{\partial \bm{r}_n}
\end{equation}
This force is non-zero only for Raman-active modes. The total Raman force is obtained by 
\begin{equation}\label{eq:Raman force}
	\bm{F} = \sum_n w_n \bm{F}_n
\end{equation}
where $w_n$ is the mode strength excited by the pump. The transition path $Q_{\rm trans}$ can be decomposed into phonon normal modes. For SnSe, MIR excitation induces forces along each $A_g$ phonon mode. The sign of the mode strength wn is determined by the differential polarizability, as calculated and depicted in Supplementary Fig.~\ref{fig:Force_direction}. The amplitude of this coefficient is derived from experimentally measured $A_g$ mode strengths presented in Fig.~\ref{fig:2}. While theoretical calculations can predict the sign of the mode strength, they lack precision in capturing the amplitude, necessitating experimental validation.

\section{Further discussion on optomechanical mechanism\label{sn:Optomechanics_mechanism}}
The optomechanical forces share the same microscopic origin as virtual or nonresonant Raman processes, \textit{e.g.}, impulsive stimulated Raman scattering (ISRS). Our previous work~\cite{Shi2025} represents the first experimental demonstration that ISRS mediated by the MIR excitation can be harnessed to drive phase transitions despite decades of efforts. The optomechanical forces described here introduce a new paradigm of phase control beyond ISRS, driving nonperturbative atomic displacements along multiple phonon coordinates. Crucially, by tuning the excitation frequencies to the positive or negative refractive index contrast between the initial and final phases, the relative atomic motions can be precisely directed toward or away from the desired state.

\section{Details on time-dependent MIR-driven atomic displacement simulation\label{sn:displacement_simulation}}
$U(\xi)$ is the equilibrium potential that depends on the position of Se atom. It can be written as \begin{equation} \label{eq:U_int}
	\begin{aligned}
		U&=a_1(\xi-\sigma)^4+a_2(\xi-\sigma)^2+a_3(\xi-\sigma)+a_4\\
	\end{aligned}
\end{equation}
where $\sigma$ is half of the distance between two local minimum in the equilibrium potential. $a_n$ are the coefficients and obtained by fitting the equilibrium potential in Fig.~\ref{fig:2}a with equation \eqref{eq:U_int}. The $\xi$-dependent dielectric constant is given by
\begin{equation} \label{eq:epsilon}
	\begin{aligned}
		\varepsilon(\xi)&=b_1\xi^4+b_2\xi^3+b_3\xi^2+b_4\xi+b_5\\
	\end{aligned}
\end{equation}
$b_n$ are coefficients by interpolating the data from first-principle calculations and then fitting with forth-order polynomials. 

The first-principle calculations yield $\varepsilon(\xi_i)$ at Se coordinates $\xi_i = (0, 0.16, 0.32, 0.48, 0.64, 0.8)\,\mathrm{\AA}$ to be $(11.3, 12.0, 18.9, 25.5, 40.8, 51.06)\,\varepsilon_0$, respectively. A phenomenological damping of $5\times10^{-13}$ s was used in solving the equation of motion. The MIR field applied is given by,
\begin{equation} \label{eq:E}
	\begin{aligned}
		F(t)=F^0 \exp[-\frac{t^2}{2\tau^2}] \exp[-i\omega t]
	\end{aligned}
\end{equation}
We can see that the ISRS formulation in Ref~\cite{Merlin1997} is a special case when we only consider $U$ to be quadratic dependent on $\xi$ (\textit{i.e.}, $a_1$=0) or when we only consider a Taylor expansion of the potential near the initial displacement until the second order. However, this is not the case in our system since our $U$ has the shape of biased double well (\textit{i.e.}, contains $\xi^4$ term). We also note that the transient change in the reflectivity during the pulse is not taken into account, and that no significant change in the threshold with pulse duration is observed. This simulation shows that the transition can happen with relatively short pulsed fields, thus helping to avoid dissipation as the system transits through the metallic state where the topological transition occurs on its way to $Fm\overline{3}m$.

\section{Details on APT measurements\label{sn:DC_APT_discussion}}

Determination of the electric field in APT is demanding, as it depends on the tip’s size and shape and thus varies throughout the measurement~\cite{Silaeva2014}. Therefore, it cannot be determined from the applied voltage alone. Nevertheless, knowledge about the exact field strength at which ions evaporate is important for the present analysis. We determine the local field in front of the tip from APT data alone~\cite{Tegg2024}, based on the post-ionization process, where ions can ionize further in high fields after evaporation~\cite{Kingham1982}. The probability to detect ions in a higher ionized state is a function of the field strength. Hence, the ratio between the occurrence of different ionization states (Charge State Ratio, CSR) in one measurement can be employed to approximate the field strength. Tegg et.al.~\cite{Tegg2024} then provided a method for numerical estimation of the electric field for a range of different elements. With this, a calculation of the field is achieved from experimental data only, making it even independent of the tip shape and other external parameters. As shown in Fig.~\ref{fig:4}b, our data for three different tips align well. We have analyzed a large number of tips from different solids and find a dependence of the PME on the applied electric field in all cases, with differences in the strength and slope of these dependence. All resulting curves can be fitted very well by a simple polynomial up to second order.  

\bibliographystyle{MSP}
%\bibliography{papers}

\providecommand{\noopsort}[1]{}\providecommand{\singleletter}[1]{#1}%

%--------------------fig:Dielectrics_overlaid---------------------------------

\begin{figure*}[htb!]
	\includegraphics[width=\textwidth]{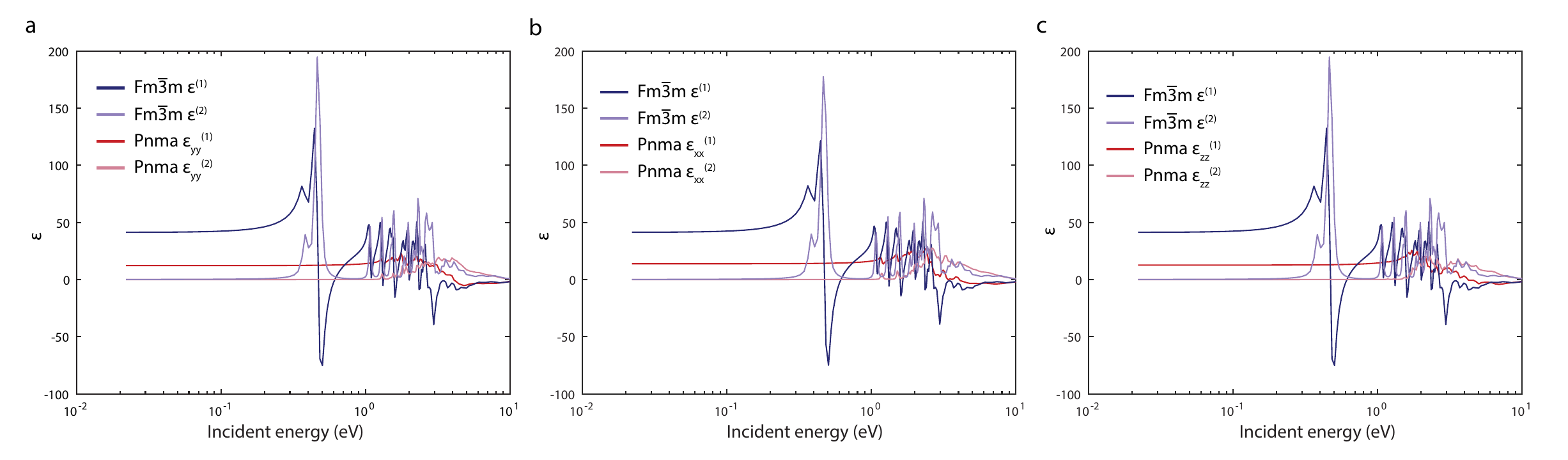}
	\centering
	\caption{\textbf{Theoretical calculated dielectric function of $Pnma$ and $Fm\overline{3}m$-SnSe.} Calculated real and imaginary part of the dielectric function of $Fm\overline{3}m$ phase and $Pnma$ phase along different crystallographic directions.}
	\label{fig:Dielectrics_overlaid}
\end{figure*}

%--------------------fig:Transient_reflectivity_dynamic---------------------------------

\begin{figure*}[htb!]
	\includegraphics[width=\textwidth]{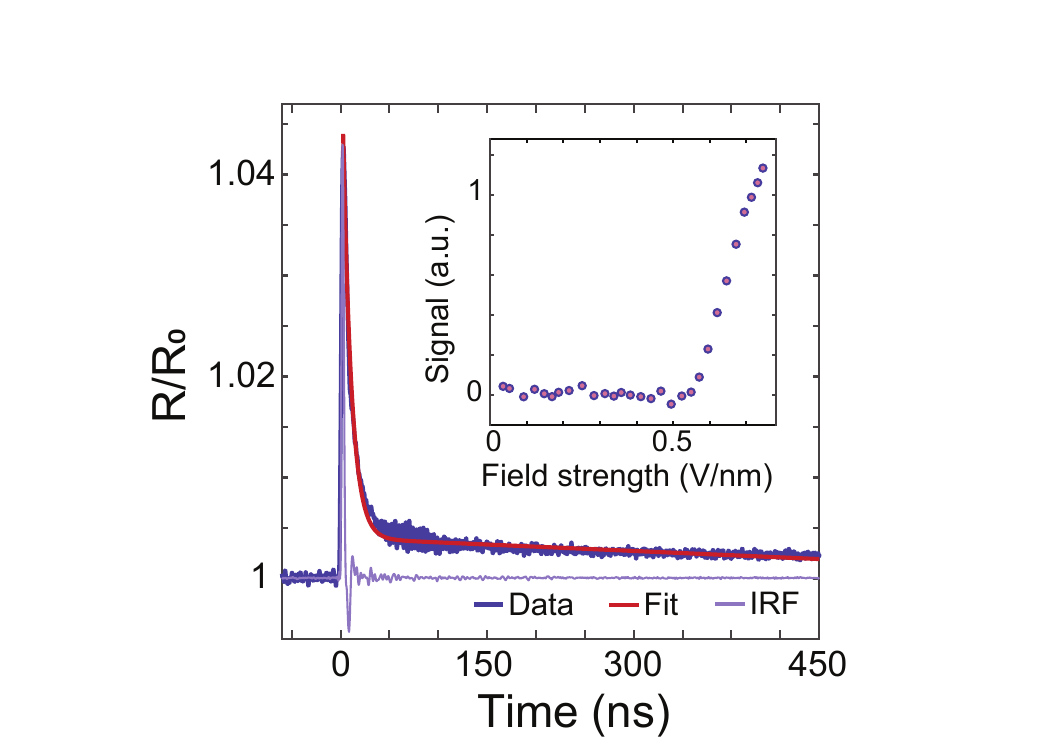}
	\centering
	\caption{Transient reflectivity dynamics at 633~nm measured with an oscilloscope show biexponential decay with long-lived responses potentially indicative of the induction of a metastable phase. IRF: instrument response function. The inset shows that the amplitude of the fast decay signal show a threshold-like field dependence.}
	\label{fig:Dielectrics_overlaid}
\end{figure*}

%--------------------fig:FTIR_SnSe---------------------------------

\begin{figure*}[htb!]
	\centering
	\includegraphics[width=0.7\textwidth]{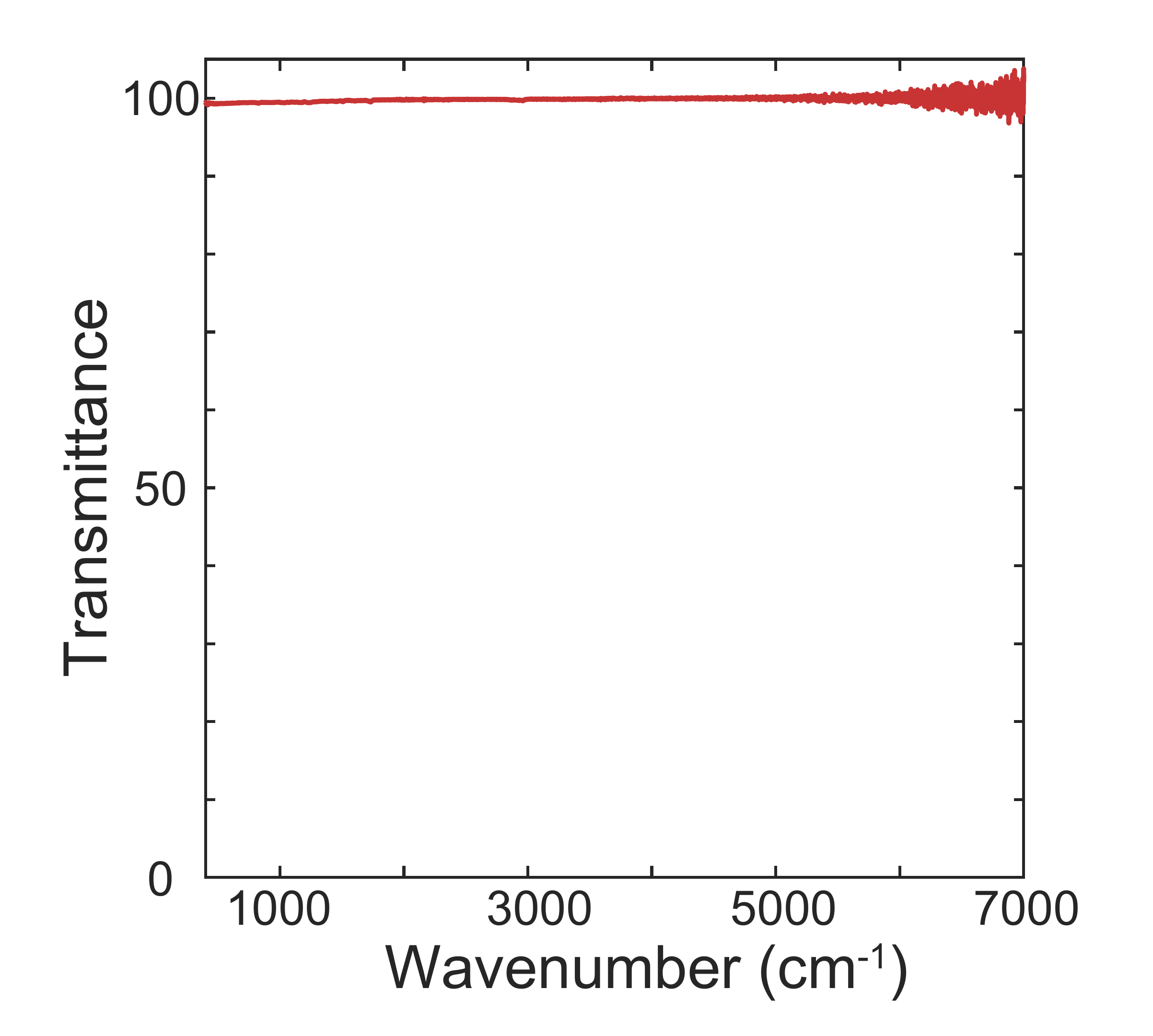}
	\caption{Fourier-transform infrared transmission measurements in attenuated total reflectance mode of bulk SnSe reveal no absorption features within the MIR range examined in this study.}
	\label{fig:FTIR_SnSe}
\end{figure*}

%--------------------fig:Imaginary_part---------------------------------

\begin{figure*}[htb!]
	\centering
	\includegraphics[width=0.7\textwidth]{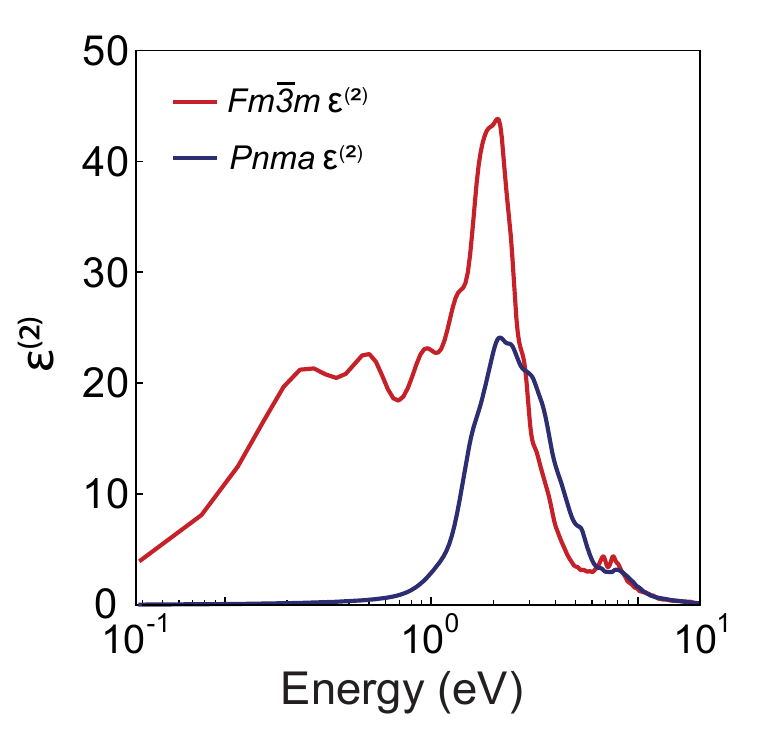}
	\caption{Theoretical calculated imaginary part of the dielectric function of $Pnma$ and $Fm\overline{3}m$-SnSe.}
	\label{fig:Imag_part}
\end{figure*}

%--------------------------------------fig:Ag_modes-----------------------------------------------

\begin{figure*}[htb!]
	\includegraphics[width=0.7\textwidth]{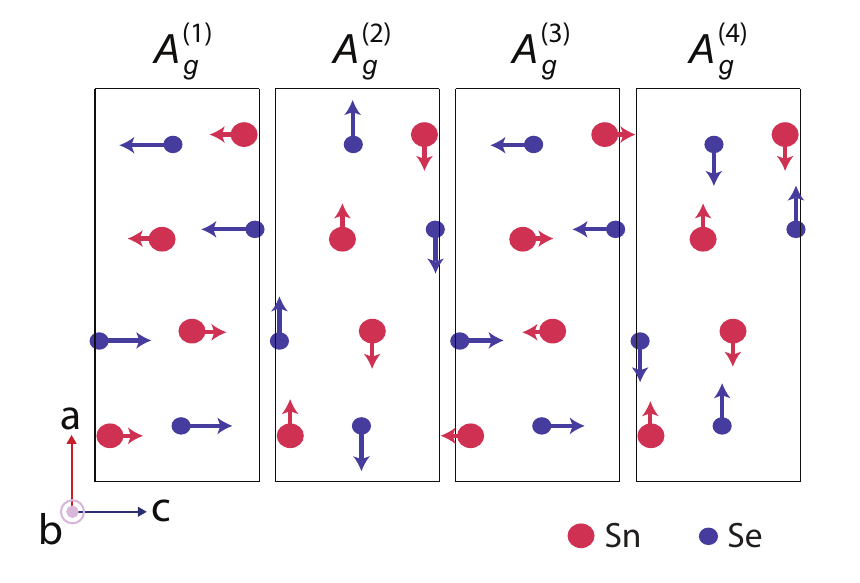}
	\centering
	\caption{\textbf{Schematic illustration of approximate atomic motions corresponding to four $A_g$ modes.}}
	\label{fig:Ag_modes}
\end{figure*}

%--------------------fig:Field_dep_offset---------------------------------

\begin{figure*}[htb!]
	\includegraphics[width=0.7\textwidth]{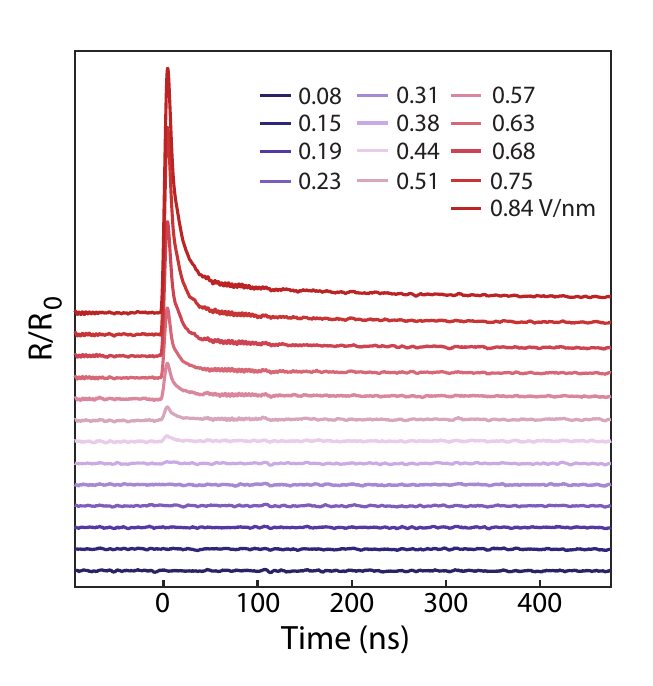}
	\centering
	\caption{\textbf{Field dependence of reflectivity change.} Reflectivity dynamics measured with an oscilloscope and a continuous-wave He-Ne laser (633 nm) as a function of MIR excitation field strengths.}
	\label{fig:Field_dep_offset}
\end{figure*}

%--------------------fig:FieldDep_Ag1---------------------------------

\begin{figure*}[htb!]
	\includegraphics[width=0.7\textwidth]{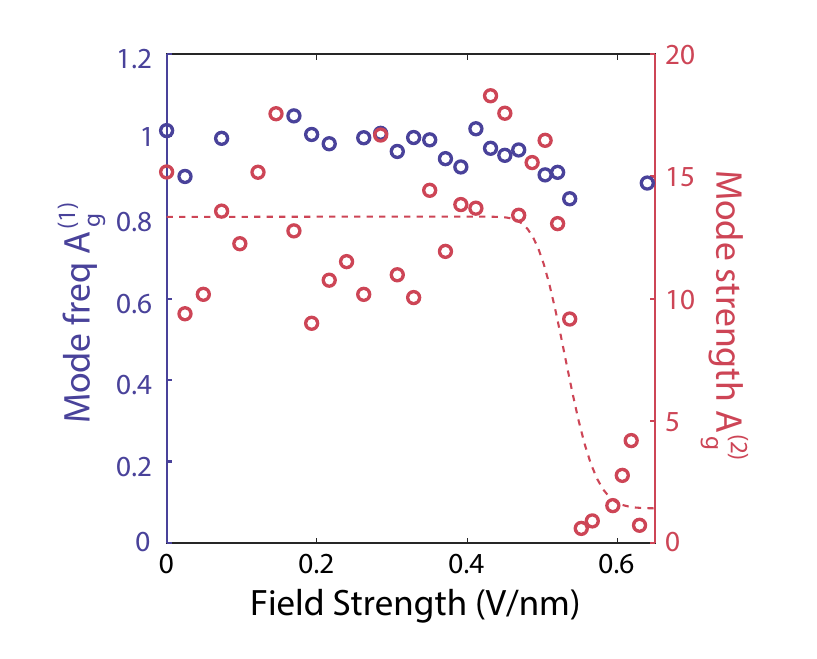}
	\centering
	\caption{\textbf{MIR field dependence scan of $A_g^{(1)}$ Raman mode.}}
	\label{fig:Fdep_Ag1}
\end{figure*}

%--------------------fig:TrRaman_delaydelay.-------------------

\begin{figure*}[htb!]
	\includegraphics[width=\textwidth]{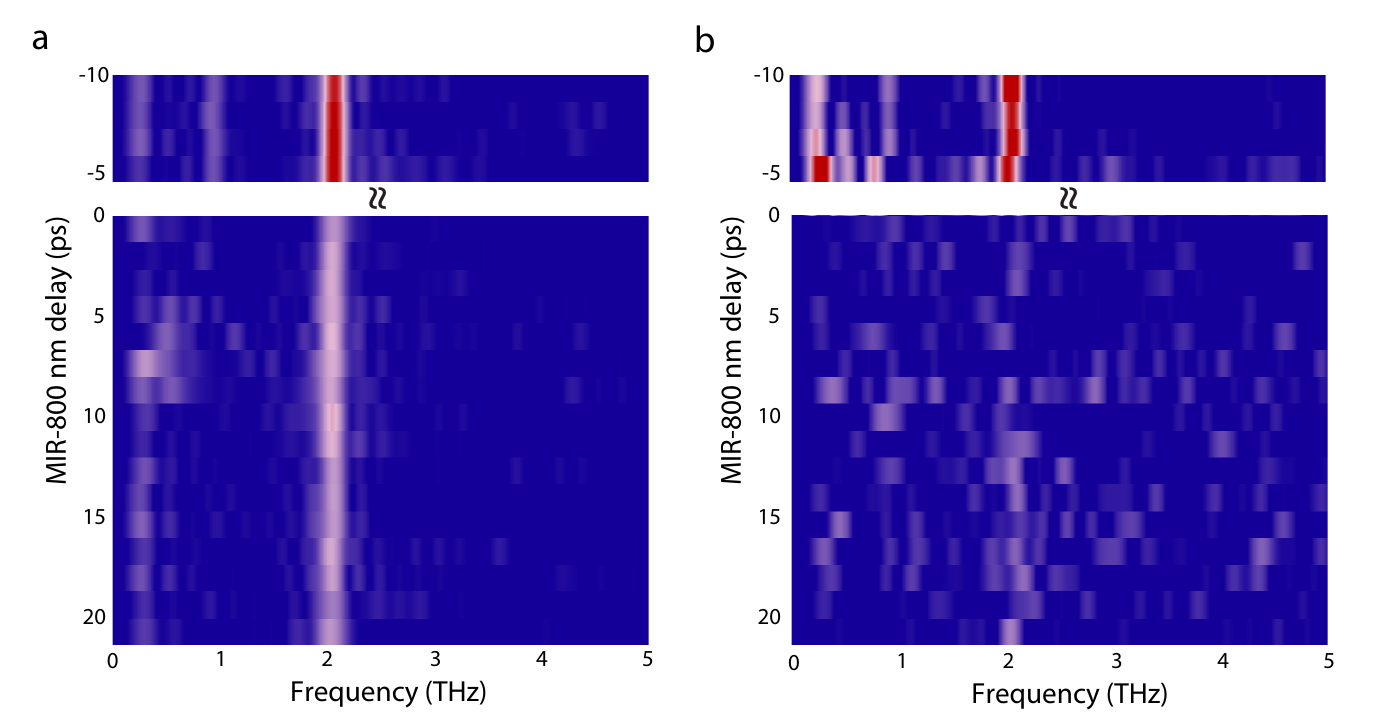}
	\caption{\textbf{Raman spectrum as a function of MIR-Raman delays.} a) Time-resolved Raman spectrum under moderate MIR field excitation at a field strength of $\sim$ 0.55~V/nm. The relative delays for Raman pump and probe pulses were scanned for $\sim$ 7 ps to resolve time-dependent Raman phonon oscillation. The MIR-Raman time delay is defined as the delay between the MIR pulse and the pump pulse of the Raman probe. Due to dramatic nonlinear responses when MIR and Raman pulses temporally overlap (shown in Supplementary Fig.~\ref{fig:Fielddelay_pp}), the phonon oscillations cannot be obtained in this time window. b) Time-resolved Raman spectrum under high MIR field excitation at a strength of $\sim$ 0.65~V/nm.}
	\label{fig:TrRaman_delaydelay}
\end{figure*}

%-------------------------fig:Fielddelay_ppexample-----------------------------
\begin{figure*}[htb!]
	\includegraphics[width=0.7\textwidth]{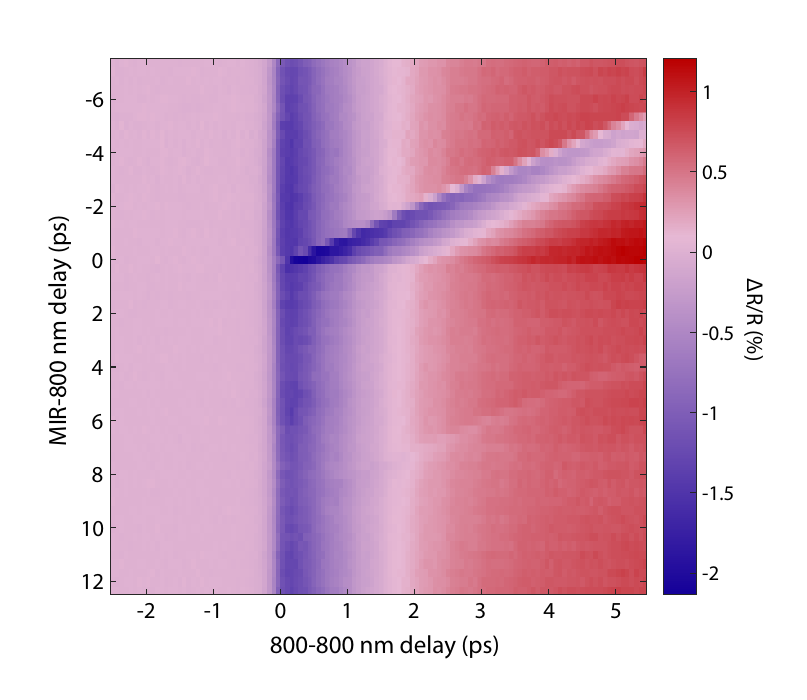}
	\centering
	\caption{\textbf{Nonlinear responses near MIR-Raman temporal overlap window.} We observe an enhanced responses near the MIR-Raman temporal overlap window, likely because a large displacement in SnSe is driven nonlinearly by intense MIR irradiation.}
	\label{fig:Fielddelay_pp}
\end{figure*}

%---------------------------fig:Temp_dep----------------------------------

\begin{figure*}[htb!]
	\includegraphics[width=\textwidth]{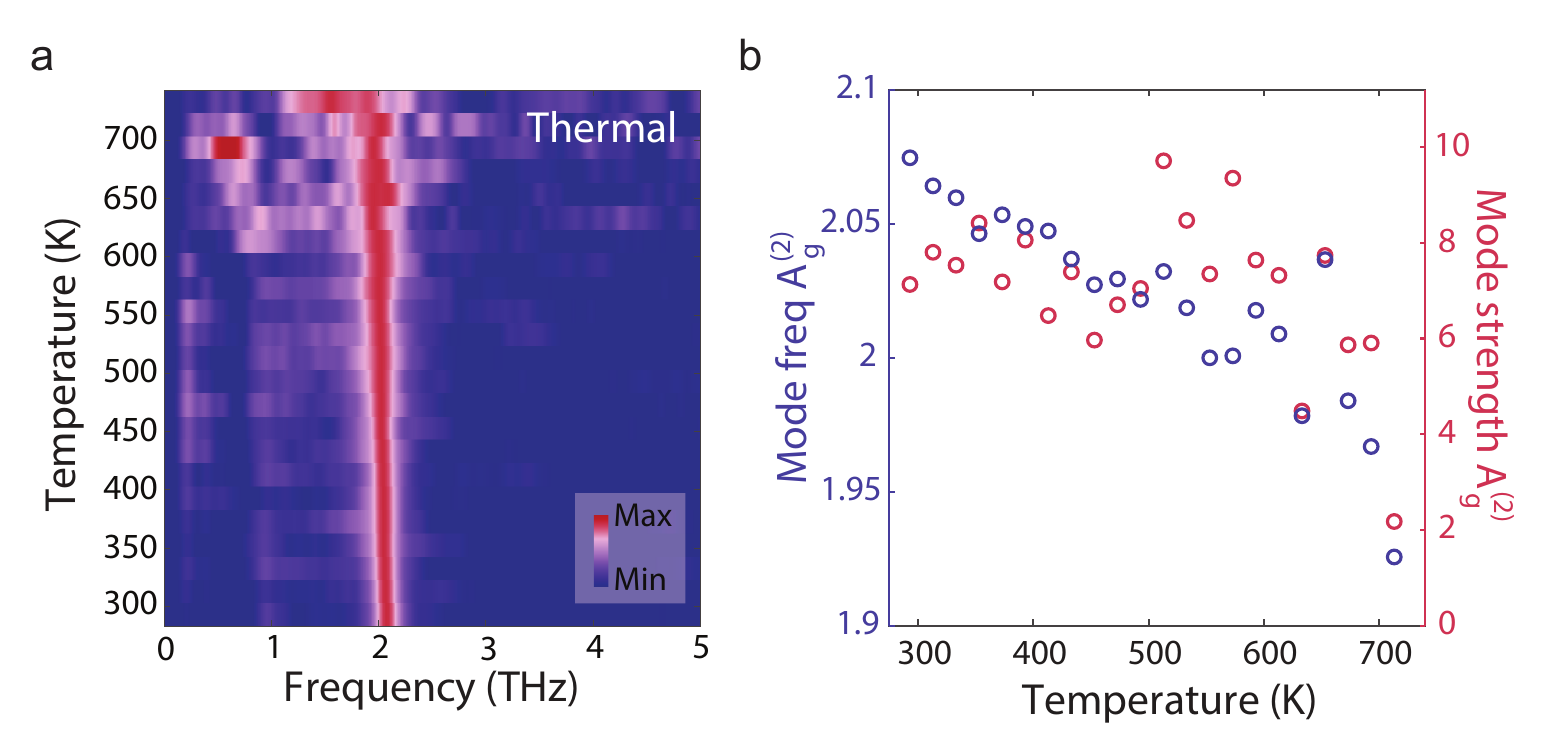}
	\caption{\textbf{Details on temperature-dependent Raman data measured with time-domain Raman scattering.} \textbf{A}, Fourier transformed reflectivity spectrum measured at evaluated temperatures. \textbf{B}, Temperature-dependent scan shows a continuous change in $A_g^{(2)}$ frequency towards phase transition temperature at around 780~K.}	
	\label{fig:Temp_dep}
\end{figure*}

%-------------------------fig:Temp_dep_analysis-----------------------------

\begin{figure*}[htb!]
	\includegraphics[width=\textwidth]{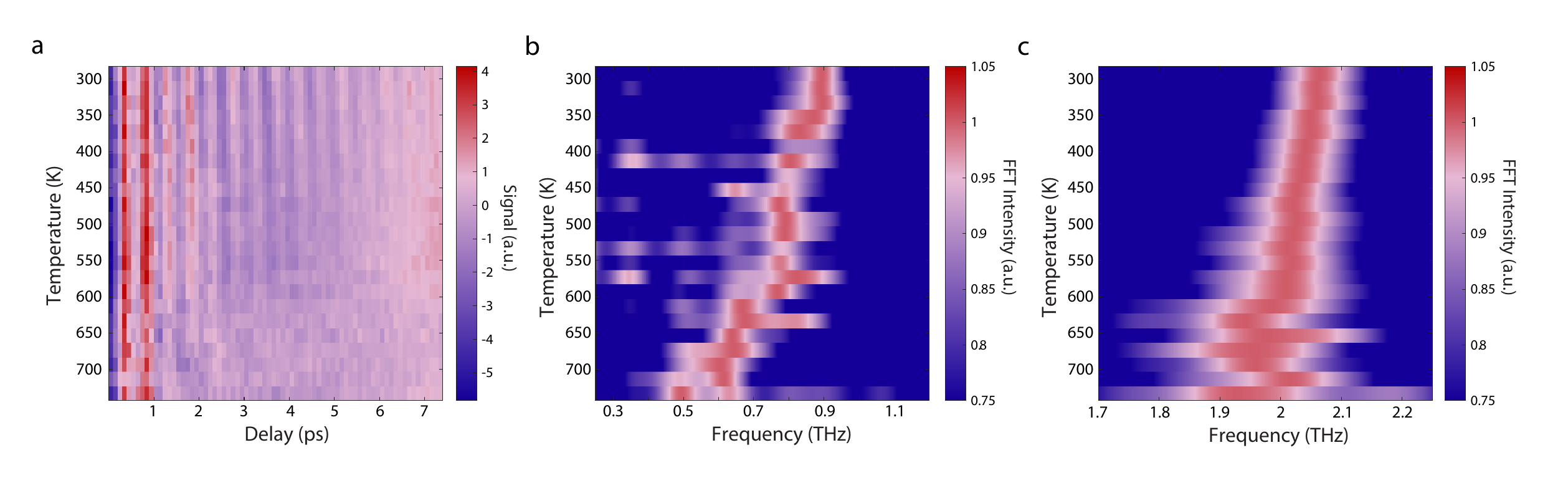}
	\centering
	\caption{\textbf{Details on temperature-dependent Raman data.} a) Time trace of phonon oscillations as a function of temperature. The background in the reflectivity change is subtracted to emphasize the phonon oscillations. b) The frequency of $A_g^{(1)}$ mode decreases at evaluated temperatures towards high-temperature $Cmcm$ phase. c) The $A_g^{(2)}$ mode softens and broadens at evaluated temperatures.}
	\label{fig:Temp_dep_details}
\end{figure*}

%-------------------------fig:Fine_FieldDep-----------------------------

\begin{figure*}[htb!]
	\includegraphics[width=\textwidth]{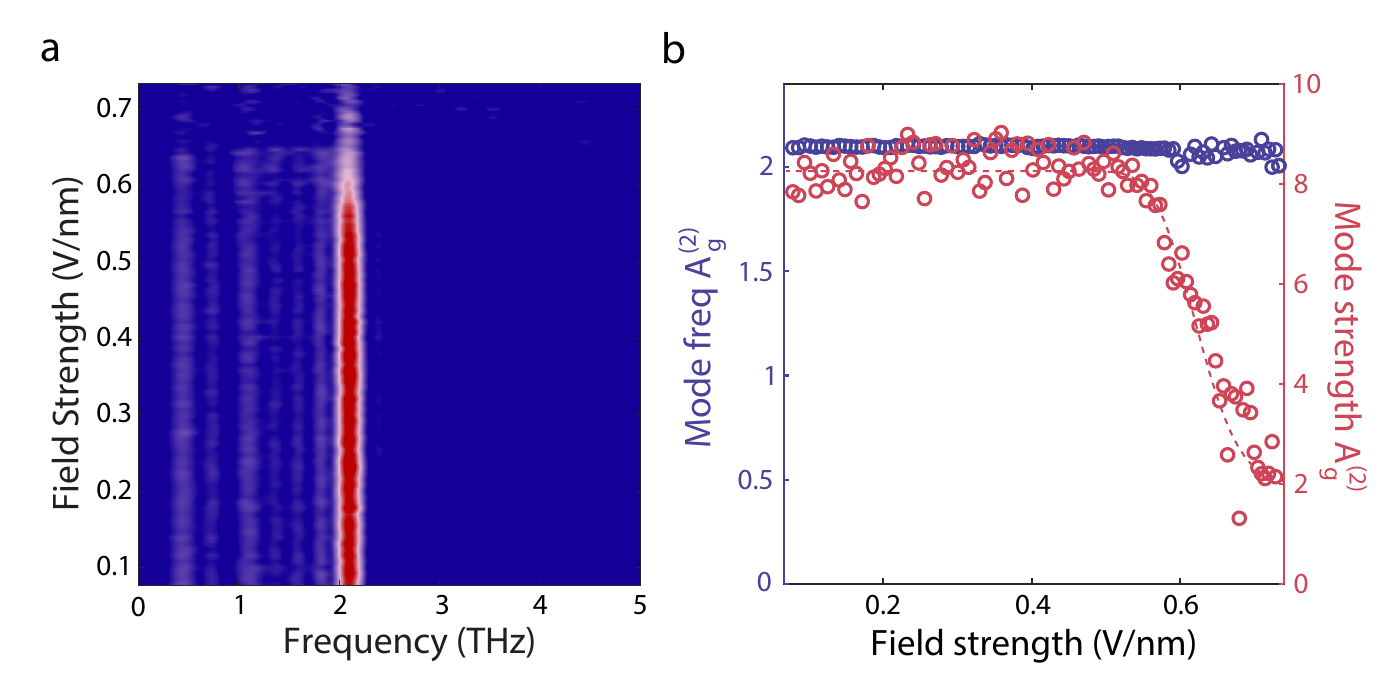}
	\caption{\textbf{Additional MIR field dependent Raman measurements.} a) MIR field dependence of Raman spectrum measured at 12 ps after MIR excitation. b) Fine field dependence scan shows an abrupt mode suppression above a critical field strength. The frequencies are extracted based on the linear prediction algorithm~\cite{Barkhuijsen1985,Led1991,Huang2022}. The same method was used for frequency extraction in Fig.~\ref{fig:3} and Supplementary Fig.~\ref{fig:Temp_dep}.}
	\label{fig:Fine_Fdep}
\end{figure*}

%-----------------fig:NIR_fluence_Raman---------------------------------
\begin{figure*}[htb!]
	\includegraphics[width=\textwidth]{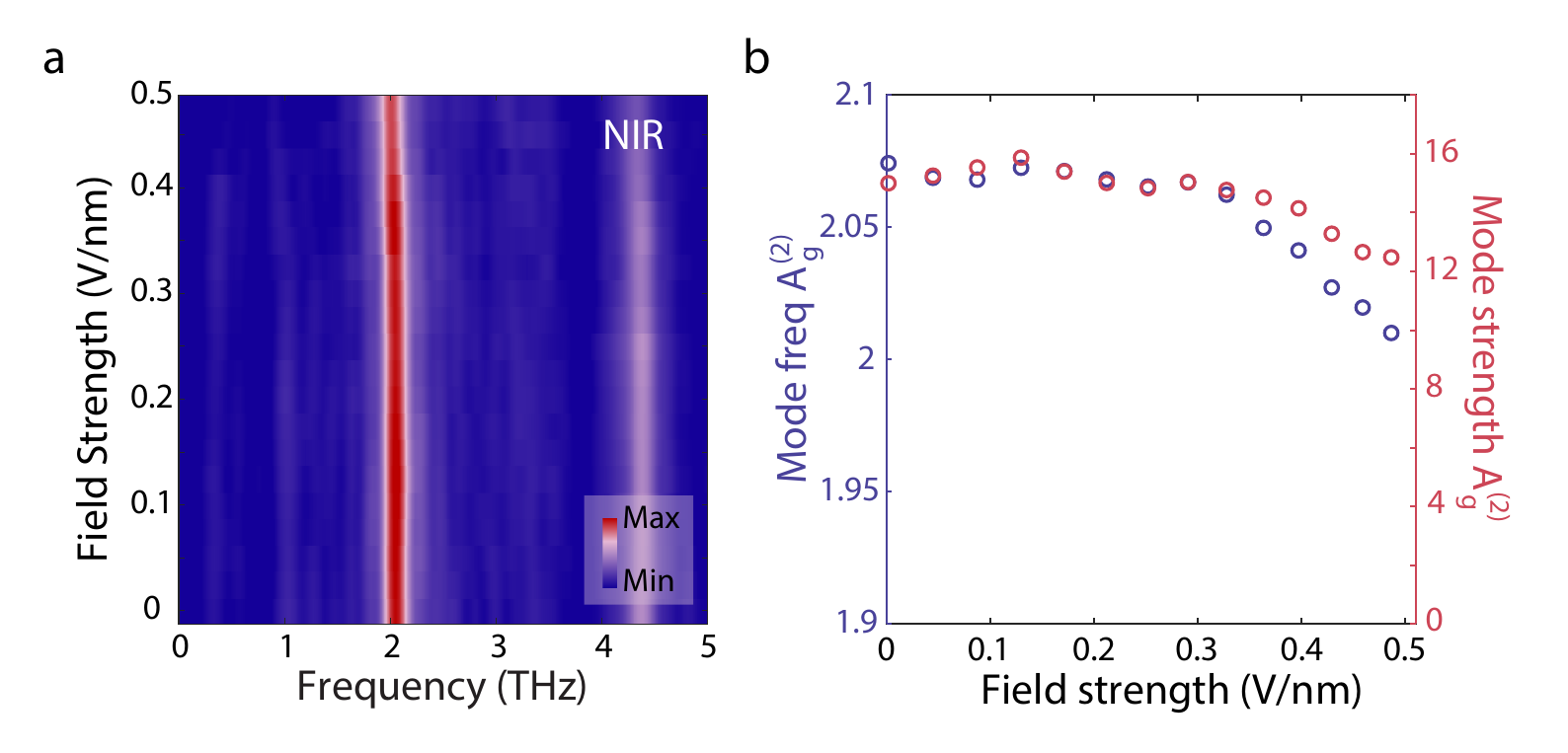}
	\caption{\textbf{NIR field dependent Raman measurements.} a) Fourier transformed reflectivity spectrum at different NIR (wavelength at 2 $\mu$m) excitation field strengths. b) Field dependence scan shows a continuous softening in $A_g^{(2)}$ frequency and a slight mode strength change under increasing NIR excitation field strengths. Note that the softening here is smaller than the heating effect. This phenomenon is probably due to the transient generation of high-density hot carriers by ultrashort NIR pulses, which may cause damage to the sample before a sufficient number of equilibrated carriers can induce phonon shifts at a level comparable to the heating.}
	\label{fig:NIR_fluence}
\end{figure*}

%----------------------------fig:MIR_TR_Fdep-----------------------

\begin{figure*}[htb!]
	\includegraphics[width=\textwidth]{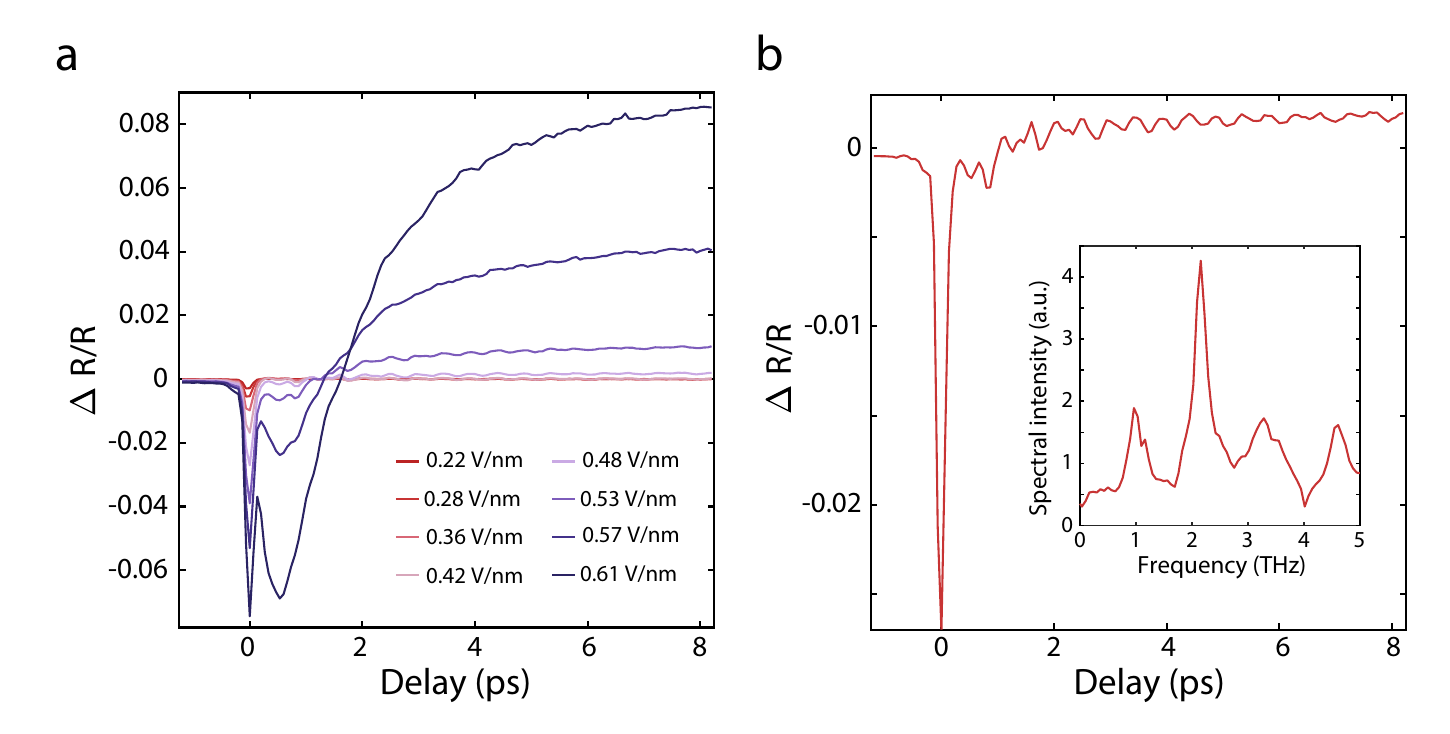}
	\caption{\textbf{Field dependence of transient reflectivity dynamics.} a) MIR (5 $\mu$m) pump 800-nm probe measurements at various MIR field strengths. At low field strengths ($\sim$ 0.22 – 0.48~V/nm), oscillations are clearly observed and correspond to $A_g$ modes. The phonon oscillations are less pronouced at field strengths higher than $\sim$ 0.53~V/nm. The emergence of a huge background signal is likely originated from a coherent acoustic signal indicative of the launch of a strain wave. b) Transient reflectivity dynamics with a low MIR field strength at 0.48~V/nm. The Fourier transform of the signal's oscillatory part is shown in the inset, and $A_g$ modes are evident.}
	\label{fig:MIR_TR_Fdep}
\end{figure*}

%----------------------------fig:Field_dep_analysis-----------------------

\begin{figure*}[htb!]
	\includegraphics[width=\textwidth]{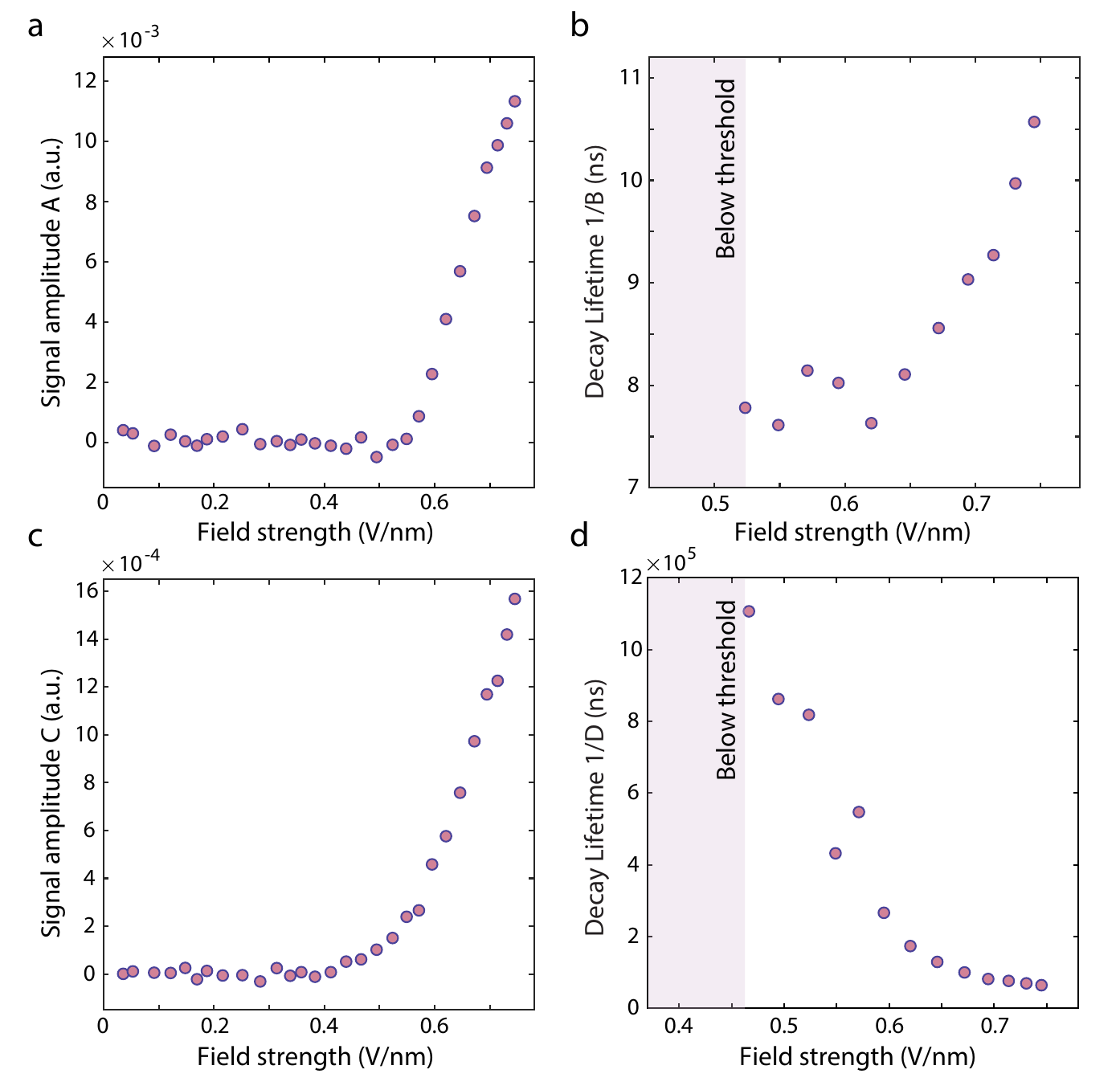}
	\centering
	\caption{\textbf{Analysis of the field dependent reflectivity.} The MIR-induced reflectivity dynamics can be fit with bi-exponential functions described by $R/R_0 = Ae^{-Bt} + Ce^{-Dt}$. The dependence of fitting parameters $A$-$D$ to the MIR field strengths is given in \textbf{a}-\textbf{d}, respectively.}
	\label{fig:Fdep_analysis}
\end{figure*}

%----------------------------fig:Fatigue_testing-----------------------

\begin{figure*}[htb!]
	\includegraphics[width=0.7\textwidth]{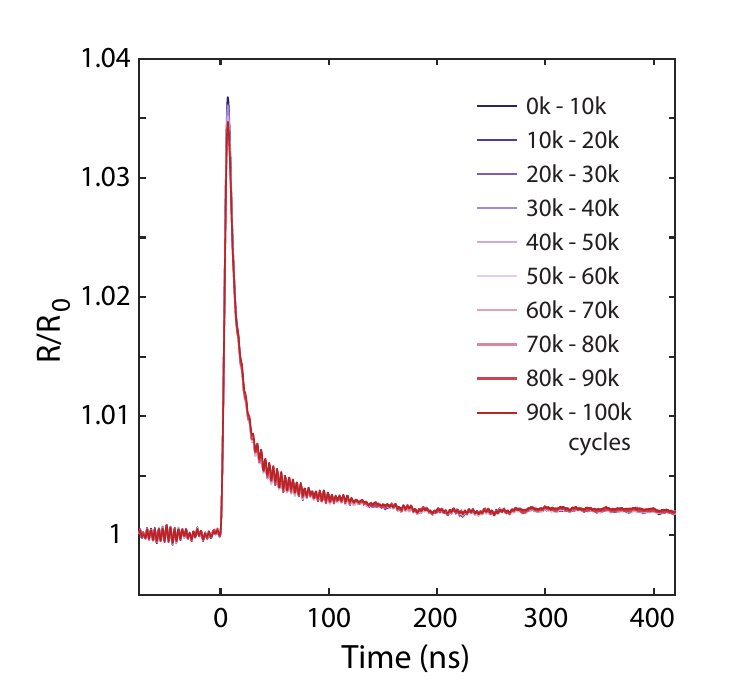}
	\centering
	\caption{\textbf{Fatigue testing under MIR excitation.} Transient reflectivity traces at 633 nm measured with an oscilloscope accumulated every 10k pulses/cycles. Ten traces shown here are acquired by averaging subsequent 10k pulses. There is only a slight decrease in the reflectivity enhancement after 100k MIR pulses, which is consistent with the martensitic and displacive type of structural distortion predicted by theory.}
	\label{fig:Fatigue}
\end{figure*}

%----------------------------fig:Theory_orbital-----------------------

\begin{figure*}[htb!]
	\includegraphics[width=\textwidth]{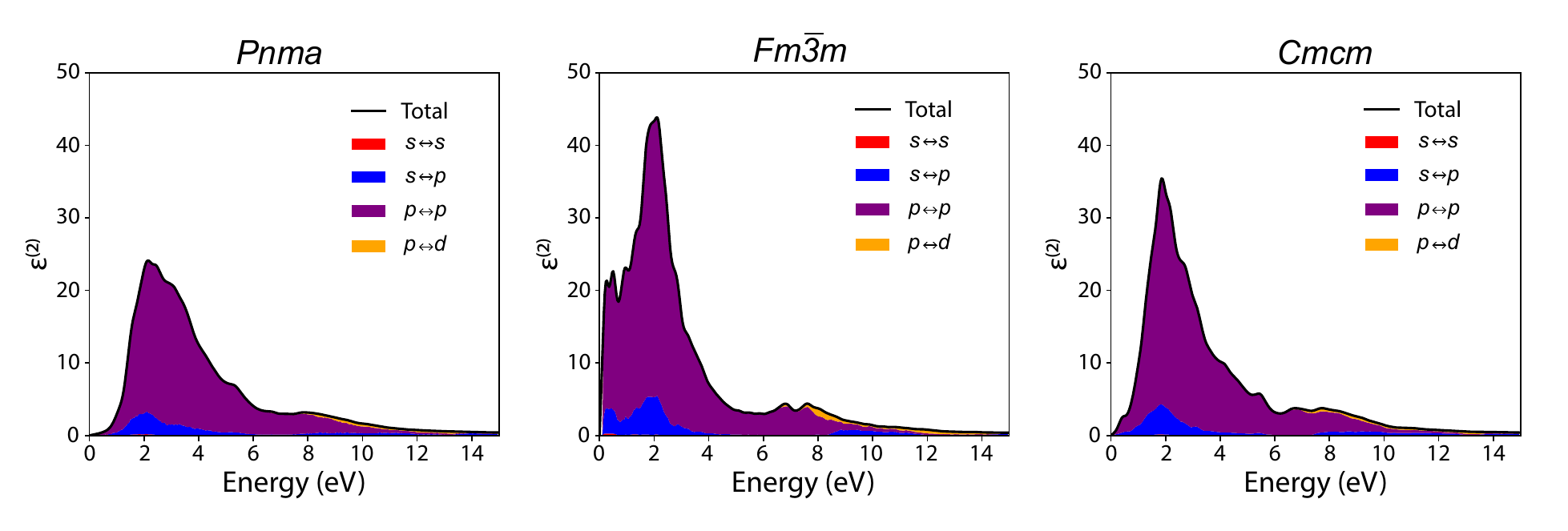}
	\centering
	\caption{\textbf{Orbital contributions of dielectric functions in three phases.} The $Fm\overline{3}m$ phase is characterized by a significantly more pronounced optical response, governed by $p$-$p$ transitions.}
	\label{fig:Theory_orbital}
\end{figure*}

%----------------------------------fig:TrRaman_setup-------------------------

%----------------------------------fig:Pol_dep-------------------------

\begin{figure*}[htb!]
	\includegraphics[width=\textwidth]{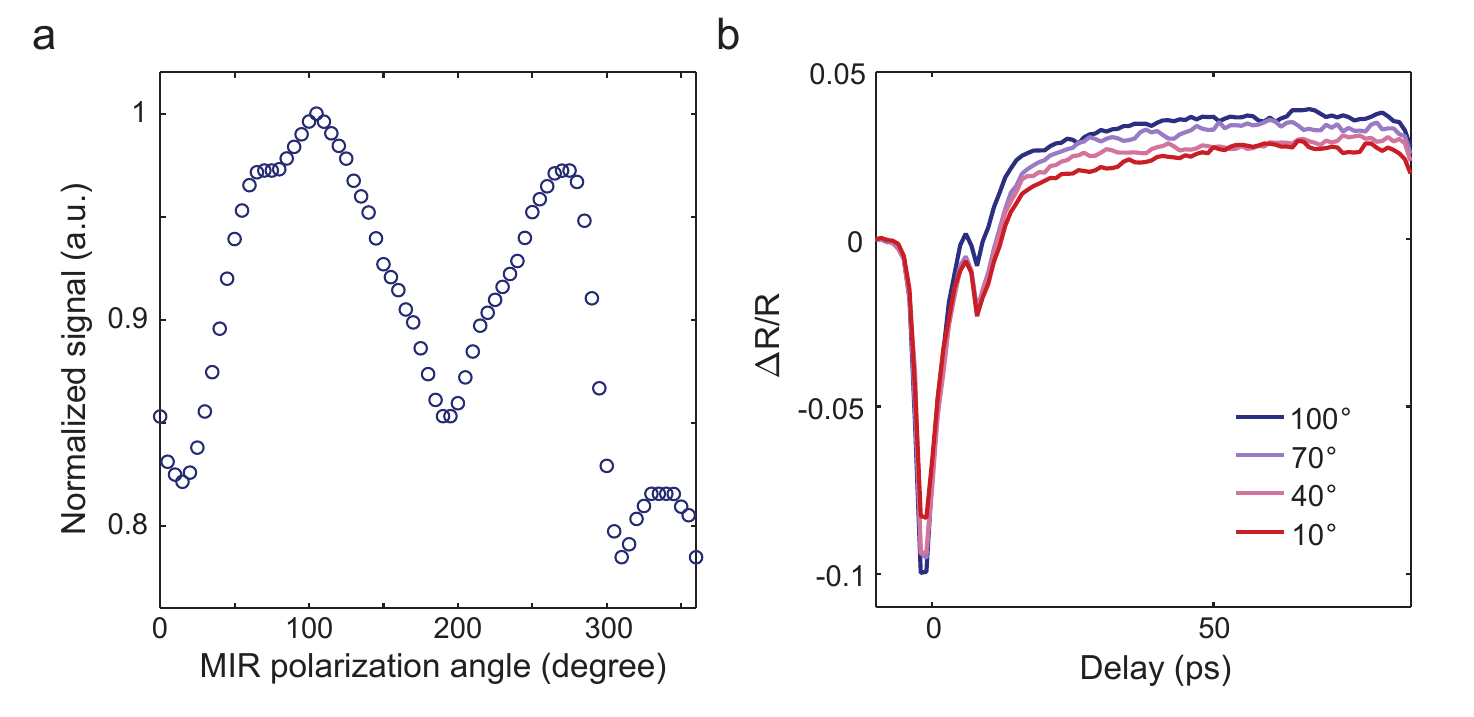}
	\centering
	\caption{\textbf{Excitation polarization dependence of MIR-driven responses.} a) Polarization angle dependence of the MIR-induced normalized reflectivity changes measured at 30 ps after time zero, which shows exhibit a two-fold symmetry and aligns with the crystal structure. The excitation pulses irradiate the sample at normal incidence. b) MIR-pump transient reflectivity changes at varying MIR incident polarization angles. Since the driving force is dominated by the refractive index difference between the initial and the final phase and because the refractive index of the final phase is so much higher than the initial phase, this polarization response is somewhat subtle.} 
	\label{fig:Pol_dep}
\end{figure*}

%----------------------------------fig:Incident_transmission_plot-------------------------
\clearpage
\newpage
\begin{figure*}[htb!]
	\includegraphics[width=0.6\textwidth]{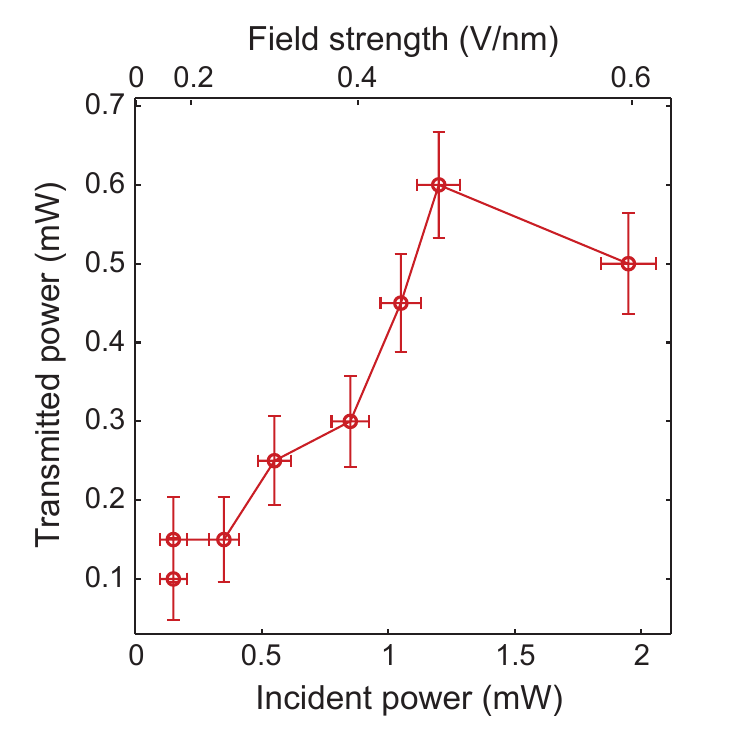}
	\centering
	\caption{\textbf{The transmitted MIR power with and without the sample at varying incident fluences.} The sample thickness is about 200 $\mu$m. Since the reflected power is challenging to measure due to the sample surface inhomogeneity, the transmitted power can be used to estimate the upper limit of the absorption. A sudden drop in transmitted power is observed at high incident power above the transition threshold.} 
	\label{fig:Incident_transmission}
\end{figure*}

%----------------------------------fig:SnSe_potential-----------------------------------------------

\begin{figure*}[htb!]
	\includegraphics[width=\textwidth]{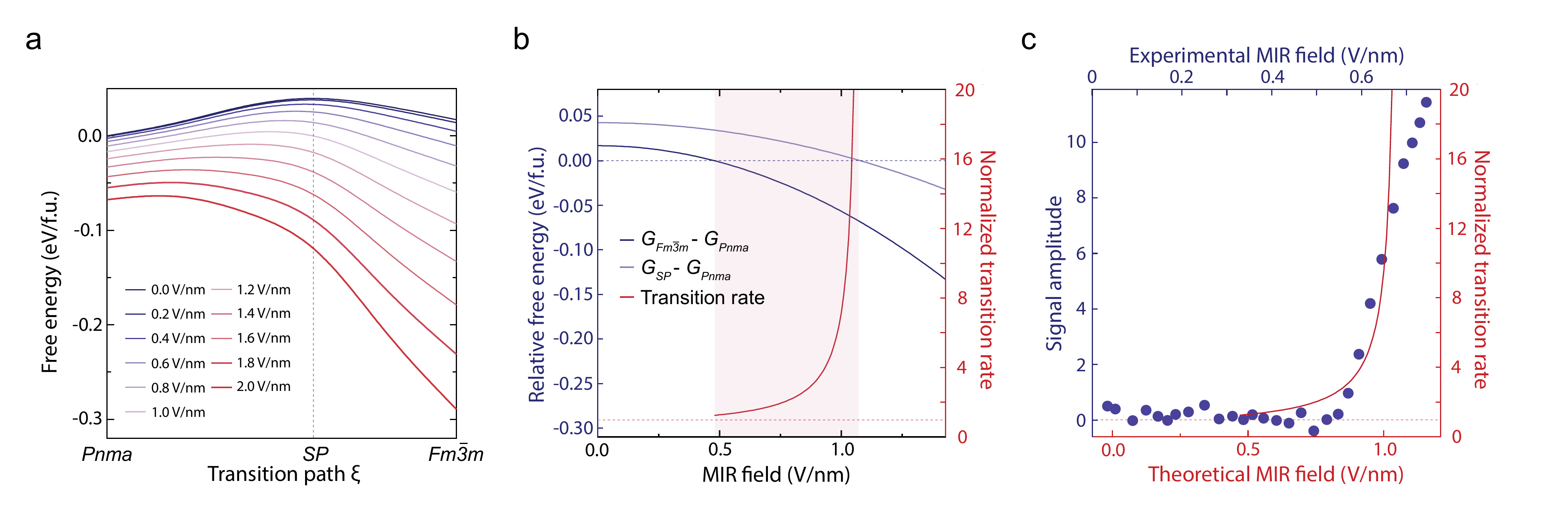}
	\caption{\textbf{Theoretical calculations of MIR-excited SnSe.} a) Grand potential of SnSe per formula unit under varying MIR fields. The energy is shifted with respect to the intrinsic (no MIR irradiation) $Pnma$ phase. SP: saddle point. b) Relative free energy and normalized transition rate as a function of MIR fields. $Fm\overline{3}m$ phase is energetically favored during the presence of a MIR field over $\sim$ 0.5~V/nm. The transition becomes barrierless when the field strength exceeds $\sim$ 1.1~V/nm. The $Pnma$-$Fm\overline{3}m$ transition rate is normalized to its light-free state and is greatly accelerated with an increasing field from 0.5 to 1.1~V/nm. c) The overlaid plot of the theoretical transition rate and the experimental results (Fig.~\ref{fig:4}) shows a qualitative agreement.}
	\label{fig:SnSe_potential}
\end{figure*}

%----------------------------------fig:SnSe_potential-----------------------------------------------

\begin{figure*}[htb!]
	\includegraphics[width=\textwidth]{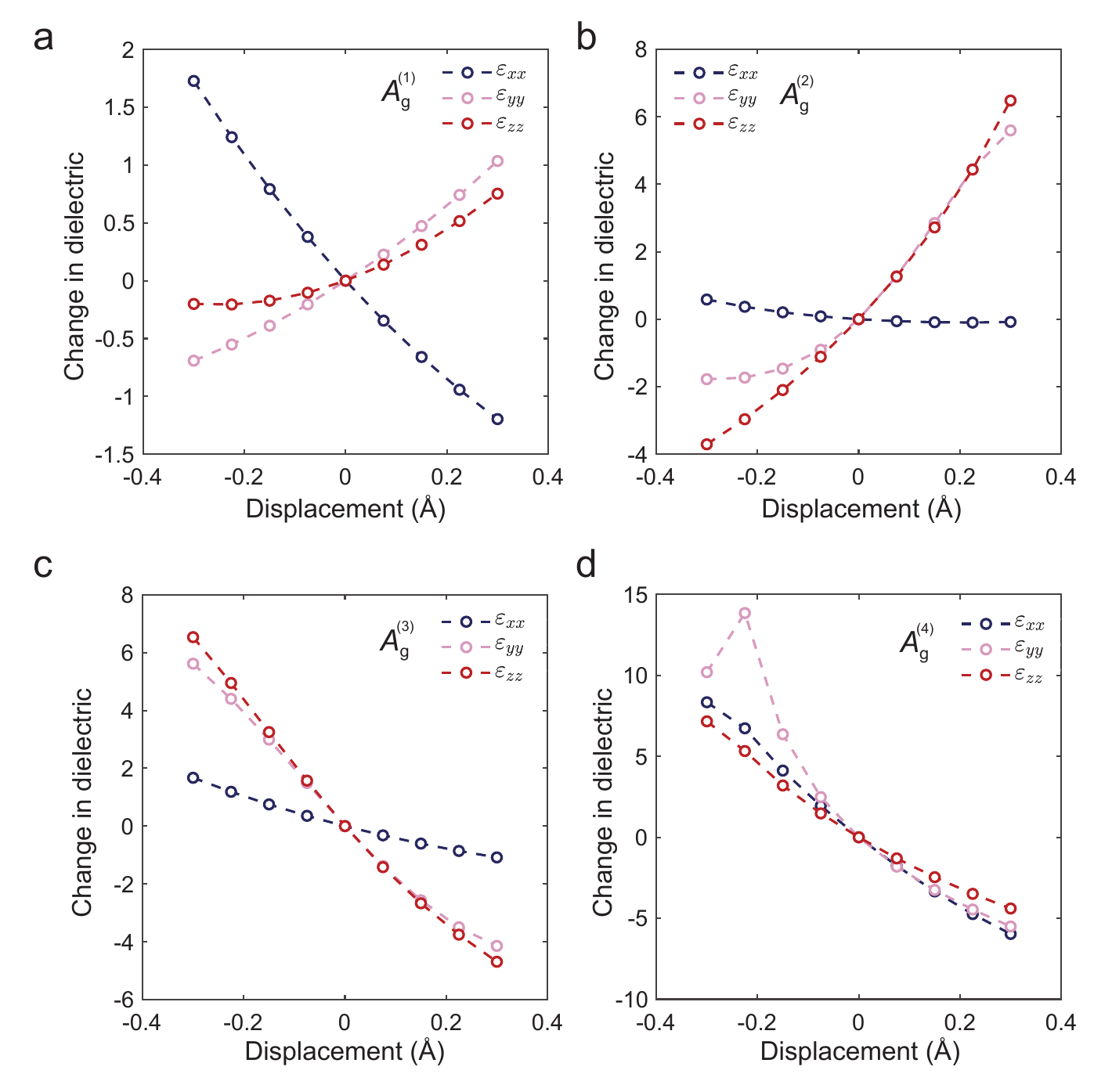}
	\caption{\textbf{Theoretical calculation of differential polarizability in SnSe.} Differential polarizability tensor ($\varepsilon_{xx}$, $\varepsilon_{yy}$ and $\varepsilon_{zz}$) in SnSe along different phonon modes ($A_g^{(1)}$,$A_g^{(2)}$,$A_g^{(3)}$, and $A_g^{(4)}$).}
	\label{fig:Force_direction}
\end{figure*}

%----------------------------------fig:PEP_Si---------------------------
\begin{figure*}[htb!]
	\includegraphics[width=\textwidth]{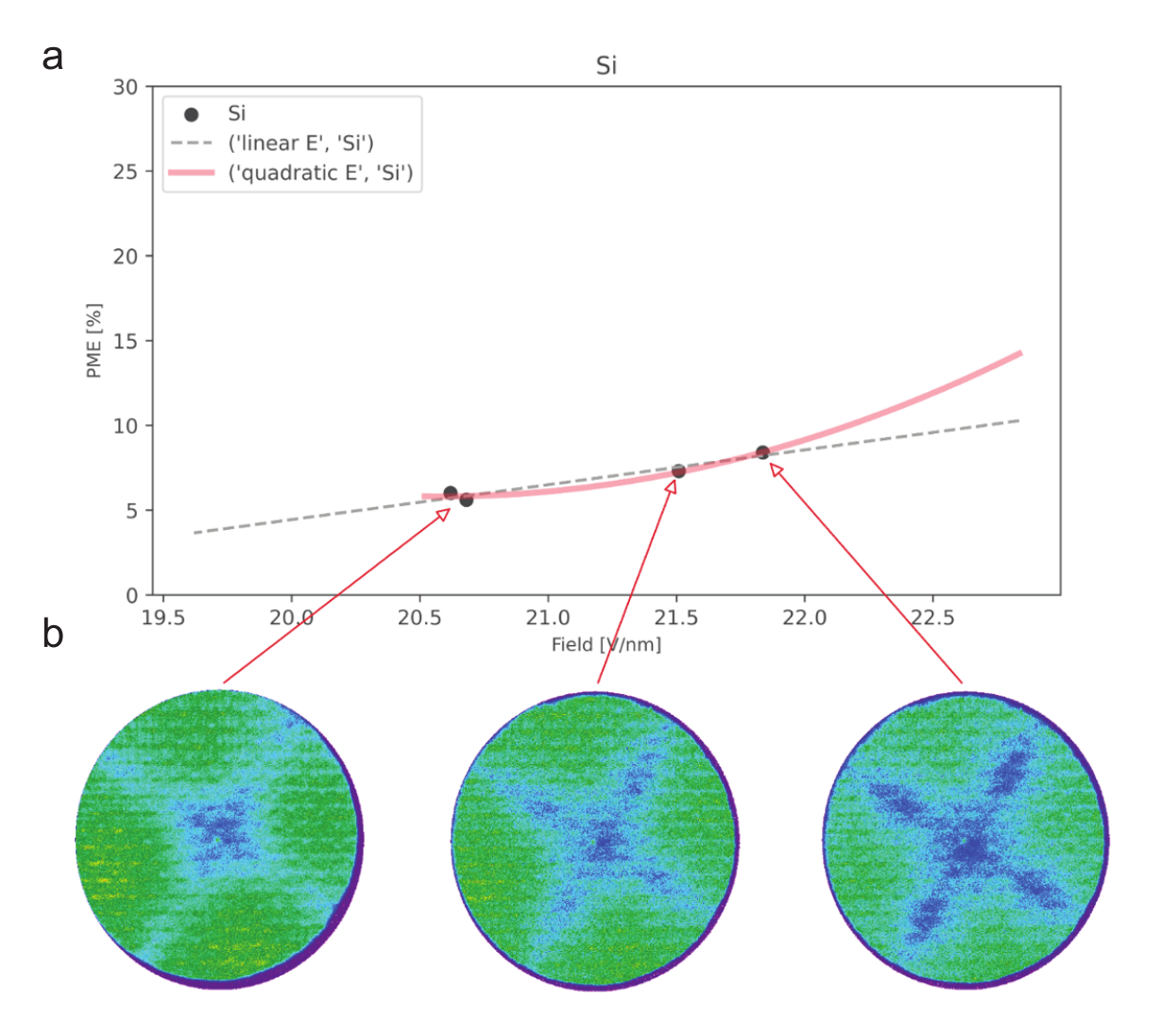}
	\caption{\textbf{PME results for silicon tip.} PME against electric field for silicon tip with corresponding detector event histograms. The PME shows a small dependence on the electric field, but neither its dependence on the field nor the detector histograms show a discontinuous change.}
	\label{fig:PEP_Si}
\end{figure*}

%----------------------------------table:Wyckoff-----------------------------------------------

\begin{table*}[htb!]
	\caption{\textbf{Wyckoff sites of $Pnma$ SnSe, and the higher symmetry structures $Cmcm$, $Fm\overline{3}m$. $x_s$ and $z_s$ are expressed in fractional coordinates \cite{Wyckoff}.}}
	\label{table:Wyckoff}
	\centering
	
	\begin{threeparttable}
		\begin{tabular}{c c c c}
			\hline
			$Pnma$ 4c &
			$Cmcm$ 4c &
			$Fm\overline{3}m$ 2d(Se) &
			$Fm\overline{3}m$ 2b(Sn) \\
			\hline
			$x_s,\frac{1}{4},z_s$ &
			$x_s,\frac{1}{4},0$ &
			$-\frac{1}{8},-\frac{3}{4},\frac{1}{4}$ &
			$-\frac{7}{8},-\frac{3}{4},\frac{1}{4}$ \\
			
			$\frac{1}{2}-x_s,\frac{3}{4},\frac{1}{2}+z_s$ &
			$\frac{1}{2}-x_s,\frac{3}{4},\frac{1}{2}$ &
			$-\frac{5}{8},-\frac{1}{4},\frac{3}{4}$ &
			$-\frac{1}{8},-\frac{1}{4},\frac{3}{4}$ \\
			
			$-x_s,\frac{3}{4},-z_s$ &
			$-x_s,\frac{3}{4},0$ &
			$-\frac{7}{8},-\frac{1}{4},\frac{3}{4}$ &
			$-\frac{1}{8},-\frac{1}{4},\frac{3}{4}$ \\
			
			$\frac{1}{2}+x_s,\frac{1}{4},\frac{1}{2}-z_s$ &
			$\frac{1}{2}+x_s,\frac{1}{4},\frac{1}{2}$ &
			$-\frac{5}{8},-\frac{3}{4},\frac{1}{4}$ &
			$-\frac{3}{8},-\frac{3}{4},\frac{1}{4}$ \\
			\hline
		\end{tabular}
	\end{threeparttable}
	
\end{table*}

%----------------------------------table:Energy_comparison-----------------------------------------------

\begin{table*}[t]
	\caption{\textbf{Energy density and lifetime comparison of light-induced transient states.}}
	\label{table:Energy_comparison}
	\centering
	
	\begin{threeparttable}
		\small
		\begin{tabular}{llllllll}
			\hline
			Category & Material & Pump & Lifetime & Length Scale &
			Heat Load$^{a}$ & Nonvolatile & Reference \\
			\hline
			
			Lattice & SnSe      & MIR     & $\sim0.2$ ms   & $3\times10^{-3}$ &
			$10^{-3}$--$10^{-4}$ & No  & This work \\
			
			Lattice & LiNbO$_3$ & MIR     & $\sim100$ fs   & $50~\mu$m &
			$8\times10^{-3}$ & Yes & Ref.~\cite{Henstridge2022} \\
			
			Lattice & GeSb      & Visible & $\infty$       & $50$ nm &
			$4$ & Yes & Ref.~\cite{Zalden2019} \\
			
			Lattice & WTe$_2$   & THz     & $>100$ ns      & $30$ nm &
			$2$ & Yes & Ref.~\cite{Sie2019} \\
			
			Lattice & SrTiO$_3$ & THz     & $\sim10$ ps    & $10$--$100$ nm &
			$0.1$ & Yes & Ref.~\cite{Li2019} \\
			
			Lattice & SrTiO$_3$ & MIR     & $>1$ min       & $10$--$100$ nm &
			$20$ & Yes & Ref.~\cite{Nova2019} \\
			
			Lattice & VO$_2$    & THz     & ns--$\mu$s     & $0.1$--$1~\mu$m &
			$0.4$ & Yes & Ref.~\cite{Liu2012} \\
			
			Lattice & LCMO      & Visible & $\infty$       & $500$ nm &
			$4\times10^{-2}$ & Yes & Ref.~\cite{Zhang2016} \\
			
			Lattice & PTO/STO   & Visible & $\infty$       & $100$ nm &
			$3$ & Yes & Ref.~\cite{Stoica2019} \\
			
			Charge & TaS$_2$    & Visible & $\infty$       & $20$ nm &
			$0.5$ & Yes & Ref.~\cite{Stojchevska2014} \\
			
			Charge & LaTe$_3$   & Visible & ns             & $44$ nm &
			$0.3$ & Yes & Ref.~\cite{Kogar2020} \\
			
			Charge & Cuprates   & MIR     & $\sim10$ ps    & $200$ nm &
			$5\times10^{-2}$ & Yes & Ref.~\cite{Fausti2011} \\
			
			Charge & K$_3$C$_{60}$ & THz  & $\sim10$ ps    & $200$ nm &
			$4\times10^{-2}$ & Yes & Ref.~\cite{Mitrano2016} \\
			
			Spin & Skyrmion     & Visible & $\infty$       & $10$--$100$ nm &
			$3$ & Yes & Ref.~\cite{Buttner2021} \\
			
			Spin & YIG:Co       & Visible & $\infty$       & $10~\mu$m &
			$4\times10^{-3}$ & Yes & Ref.~\cite{Stupakiewicz2017} \\
			
			Spin & PCMO         & MIR     & ns--$\mu$s     & $500$ nm &
			$2\times10^{-2}$ & Yes & Ref.~\cite{Rini2007} \\
			
			Spin & TmFeO$_3$    & THz     & $<1$ ns        & $13~\mu$m &
			$8\times10^{-2}$ & Yes & Ref.~\cite{Schlauderer2019} \\
			
			\hline
		\end{tabular}
		
		\begin{tablenotes}
			\item[a] Heat load: energy applied divided by the volume of the switched material.
		\end{tablenotes}
		
	\end{threeparttable}
\end{table*}